\documentclass[12pt]{iopart}
\pdfoutput=1
\usepackage{graphicx}

\usepackage{iopams}  
 \usepackage{graphicx}     
\begin{document}

\title[Persistence in coupled logistic map]{Stretched exponential 
dynamics of coupled logistic maps on a small-world network}

\author{Ashwini V. Mahajan$^1$ and Prashant M. Gade$^2$}
\address{$^1$ Department of Physics, Savitribai Phule Pune University, Pune 411007, INDIA, $^2$ PG Department of Physics, Rashtrasant Tukadoji Maharaj Nagpur
University, Nagpur, 440033, INDIA.}
\ead{prashant.m.gade@gmail.com}
\vspace{10pt}
\begin{indented}
\item[]June 2017
\end{indented}

\begin{abstract}
We investigate the dynamic phase transition from partially or fully 
arrested state to
spatiotemporal chaos in coupled logistic maps on a small-world network. 
Persistence of local variables in coarse grained sense acts as an 
excellent order parameter to study this transition. We investigate 
the phase diagram by varying 
coupling strength  and small-world rewiring probability $p$ of 
nonlocal connections. 
The persistent region is a compact region bounded by two 
critical lines where band-merging crisis occurs. 
On one critical line, the persistent sites shows a 
nonexponential (stretched exponential)
decay for all $p$ while for another one, it shows crossover  
from nonexponential to exponential behavior as $p \rightarrow 1$.
With an effectively antiferromagnetic coupling, 
coupling to two neighbors on either side leads to exchange frustration.
Apart from exchange frustration,
 non-bipartite topology and nonlocal couplings
in a small-world networks
could be reason for anomalous relaxation.
The distribution of trap times in asymptotic regime has a long tail as well.
The dependence of temporal evolution of persistence 
on initial conditions is 
studied and a scaling form for persistence after waiting time is proposed.
We present a simple possible model for this behavior.
\end{abstract}

%
\vspace{2pc}

\section{Introduction}
The study of spatially extended systems has progressed 
very well from last few decades. 
Since such  a system is also a many body system, the entire toolbox 
of statistical mechanics is applicable to it. 
However, the properties are not so well studied from the statistical 
mechanics viewpoint. Spatiotemporal dynamical system 
is a  non-equilibrium systems 
and dynamical phases and phase transitions therein, usually (but not always) 
have no equivalent analogue in equilibrium systems. 
They neither have a Hamiltonian nor obey detailed balance.
It may not even happen that some coarse-grained 
quantities obey Langevin dynamics. Thus
a reduced free energy description may not be derived for these systems 
\cite{miller-huse}.

Coupled map lattice (CML) is one of the most popular model 
to explore the dynamics of
spatially extended systems \cite{kaneko}. It displays wide range of 
spatiotemporal patterns \cite{jabeen,gade,jalan}. 
Networks is another emerging field which has received wide attention 
in recent times and CML studies have been extended to them. 
Connectivity patterns in a wide range of systems in nature as well 
as in society are
described by complex networks. Examples include neural networks, 
food webs or communication media.  They have a complex architecture and
do not have d-dimensional lattice as underlying topology. 
Several dynamical phenomena ranging from the
spread of diseases to the spread of 
rumors or the extinction of species can occur on them. 
Random network \cite{erdos}, small-world (SW) network \cite{watts} 
and scale-free network \cite{albert} are the most popular and 
extensively studied models. They have different degree distributions and 
are useful in different contexts\cite{barabasi}. 
In past few years, we have witnessed several studies on spatially extended 
dynamical systems on complex networks \cite{review,mahajan,ashwini}. 
In SW networks, one starts with a regular network where each site 
is connected to $k$ nearest 
neighbors. Each link is
rewired randomly with probability $p$ and connected to a randomly 
chosen site on lattice. This network has a high clustering coefficient like 
that of regular networks and has a low characteristic path length like 
random networks for small $p$.  
SW networks extrapolate between regular graph for $p=0$ to random
graph for $p=1$.
Equilibrium and few non-equilibrium systems are studied on SW network. 
It has been established that equilibrium transitions on small-world network 
fall in the universality class of mean-field models for $p\ne 0$
while situation is not 
so clear in non-equilibrium systems\cite{sinha}.

We study coupled logistic maps on a SW network 
where each site is
coupled with four other sites. Studies on 
coupled logistic maps in 1-dimension 
show the transition to  an antiferromagnetic order 
for certain coupling strength and at the critical point, 
decay of persistence matches with Glauber-Ising model\cite{GGS}. 
We argue  that coupled logistic maps  have
an effective antiferromagnetic coupling since $f(x)$ is decreasing
function near nonzero fixed point. 
If we couple 
each site to the nearest as well as
next-nearest neighbors, it leads to exchange-frustration
for antiferromagnetic
coupling.
Besides, SW network 
does not even have bipartite topology
and any long-range order with effectively
antiferromagnetic couplings cannot exist.
For small coupling, we observe dynamically 
arrested states which do not have any apparent spatial order but
are frozen temporally in a coarse grained sense. Usual order parameters such 
magnetization or sublattice magnetization do not work for this state.
 We study
the temporal properties of correlations for such transition.

Frustrated systems can have anomalous relaxation properties.The 
Kohlrausch-Williams-Watts 
function (KWW or stretched exponential)\cite{kohlrausch,williams} 
is defined as :
\begin{equation}
f_{\beta}(\frac{t}{\tau_0})=\exp[-(\frac{t}{\tau_0})^\beta]
\end{equation}
for $t > 0$ and $0 < \beta < 1$ with $\tau_0 > 0$ an effective time constant, 
is usually used as a phenomenological description of relaxation in a 
variety of complex materials, disordered, frustrated systems such as 
spin-glasses, polymers, coupled oscillators, jamming transitions, driven 
temporal networks, coupled nonlinear systems, molecular systems, glass 
soft matter, porous and noncrystals silicon
\cite{glotzer,hunt,chaudhuri,mata,amritkar,simdyankin,cipelletti,pavesi}. 
In first passage times, stretched exponential behavior 
has been 
observed in 
fluctuation barrier problem\cite{taillefumier} and
currency exchange rates\cite{kurihara}. 
Such behavior is essentially a signature of long-time 
confinement of dynamics to some regions in phase space.
In these systems, frozen disorder leads to slow dynamics and various theories 
have been proposed to explain these.

We conduct a detailed study of phase transition from partially or fully
arrested state 
to spatiotemporal chaos as a dynamics phase transition. We consider local 
persistence as an appropriate order parameter to quantify this transition. 
Persistence quantifies the fraction of sites
in system which have not deviated even once from its initial state.
This quantifier is analogous to, but much stronger than Edwards-Anderson
order parameter used in studies on spin glass. These studies are
also similar to branch of statistics called `survival analysis' which
studies durations till a given event such as death in living
systems or failure in mechanical systems happens\cite{lawless}.
The most popular fitting function in this analysis is Weibull distribution
which has stretched exponential survival function for some parameters.
This survival function is expected
when hazard rate is nonstationary, it decreases
in time and there are multiple causes of failure. On the other hand,
we expect exponential survival function when hazard rate is constant and
mortality is independent of age. We also study other quantifiers such as
spin flipping probability
and trap time distribution.
Spin flip probability is analogous to (but not same as) hazard rate in 
mortality studies since spins can flip multiple times.
We also present a simple model for this behavior.


\section{Model}
\label{sec2}
We construct coupled map lattice on a small-world network.
Time evolution
of the variable
$x_i(t)$ at site $i$ at time $t$ depends on itself and 
 $4$ nearest neighbors ($1\le i\le N$ where $N$ is system size). 
We assume periodic boundary conditions and 
$x_i(0)$ are random variables in the interval $(0,1)$.
\begin{equation}
\begin{tabular}{ r l }
\(x_{i,t+1}\)&\(= (1-\epsilon)f(x_{i,t}) +
\frac{\epsilon}{4} \lbrace f(x_{\xi_1(i),t})+ f(x_{\xi_2(i),t})
+ f(x_{\xi_3(i),t}) + f(x_{\xi_4(i),t})\rbrace\)  
\end{tabular}
\end{equation}
where $\epsilon$ is the coupling strength. Using the
above definition of SW network, 
$\xi_1(i)=i+1,  \xi_2(i)=i+2, \xi_3(i)=i-1, \xi_4(i)=i-2$ chosen 
with probability $1-p$ and any randomly chosen lattice site  
with probability $p$. This topology does not change during evolution.
The topological properties such as clustering coefficient and average 
path length change with $p$. Question is how the 
dynamical properties
and dynamical transitions  are affected as we change structural properties.
 
The on-site map is logistic map $f(x) = \mu x(1-x)$.
where $\mu$ is the map parameter. We study the
system in chaotic regime by 
keeping map parameter $\mu = 3.9$ constant by 
varying $\epsilon$ and $p$ from $[0,1]$.

The nonzero fixed point of the logistic map is given by
\begin{equation}
x^* = 1-1/{\mu} 
\end{equation}

To explore the dynamical phases and transitions therein,  
we need to simulate the system for larger time over large configurations 
and an appropriate order parameter has to be defined. We consider 
local persistence as an order parameter. Persistence 
has been helpful in studying transitions to a  
dynamically arrested state \cite{mahajan,abhijeet}. It
is an analogue of first passage times. 
In spin systems, persistence $P(t)$ is number of sites which did not 
change their initial value {\em {even once}} till time $t$. 
This is a non-Markovian quantity which requires knowledge of complete
time evolution of the system for its computation.
We coarse-grain the dynamics by labeling the sites with 
$x_{i}'s$ values greater 
than $x^*$ as `up' spins and rest as `down' spins. 
For logistic map,
this definition needs to be modified 
as the slope of the map at $x^*$ is negative for $\mu>2$. 
We expect the sites with value greater than $x^*$ returns a 
lower value than $x^*$ and vice versa. 
Hence we use the generalized definition  and check
for persistence at all even time steps as in ref. \cite{GGS}. 
The spin variable $s_i(t)=1$ if $x_i(t)\ge x^*$ 
otherwise $s_i(t)=-1$. The local persistence $P(\tau)$ is defined as follows-
$P(\tau)$ is a fraction of sites for which $s_i(0)=s_i(2 \times t)$  
for all  $0\le t\le \tau$.

\section{Phase Diagram and Dynamic behavior}
\begin{figure}
\centering
\includegraphics[width=0.5\textwidth]{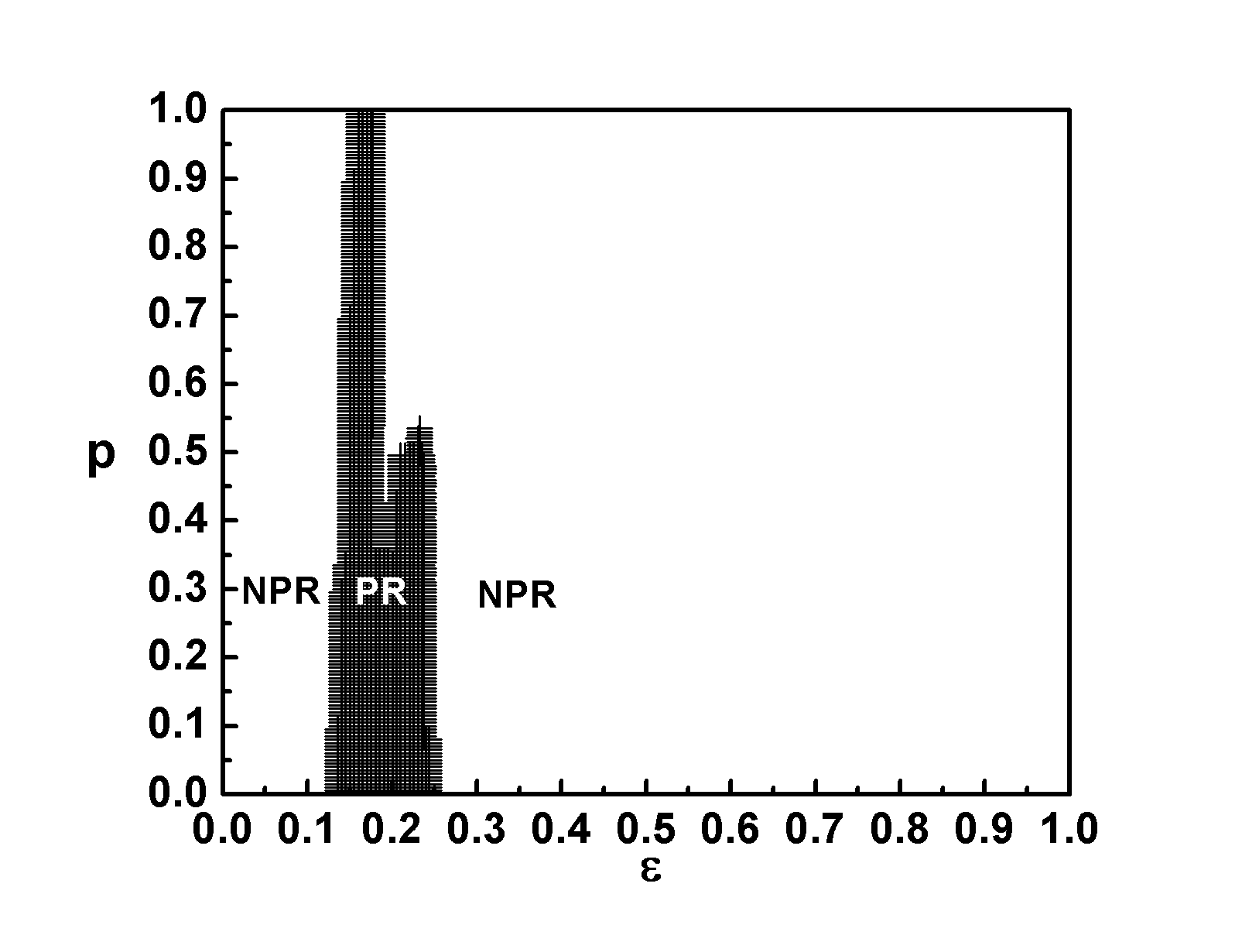}
\caption{\label{phase}Phase diagram of coupled logistic map. The dark region in figure is persistent region. The region outside dark region is non-persistent region.}
\end{figure}

We examine the system by varying parameters mentioned above. 
In $\epsilon-p$ space, a significant part 
is occupied by dynamically arrested state. They retain the memory of 
initial state 
for indefinite time leading to a non-zero asymptotic 
value of persistence. 
The persistent region ($PR$) and non-persistent region ($NPR$) are plotted in 
Fig.\ref{phase}. 
The persistent region is compact and is bounded by two critical 
lines. The dynamic behavior on the critical lines is different at 
different points. The phase diagram remains qualitatively same 
for larger values of $N$ and for longer time. This diagram is obtained 
only after waiting for very long times (for $N=10^{4}$ and 
wait for $5 X 10^7$ transient), otherwise spurious persistent points 
can appear in non-persistent region.
\begin{figure}
\centering
\includegraphics[width =0.5\textwidth]{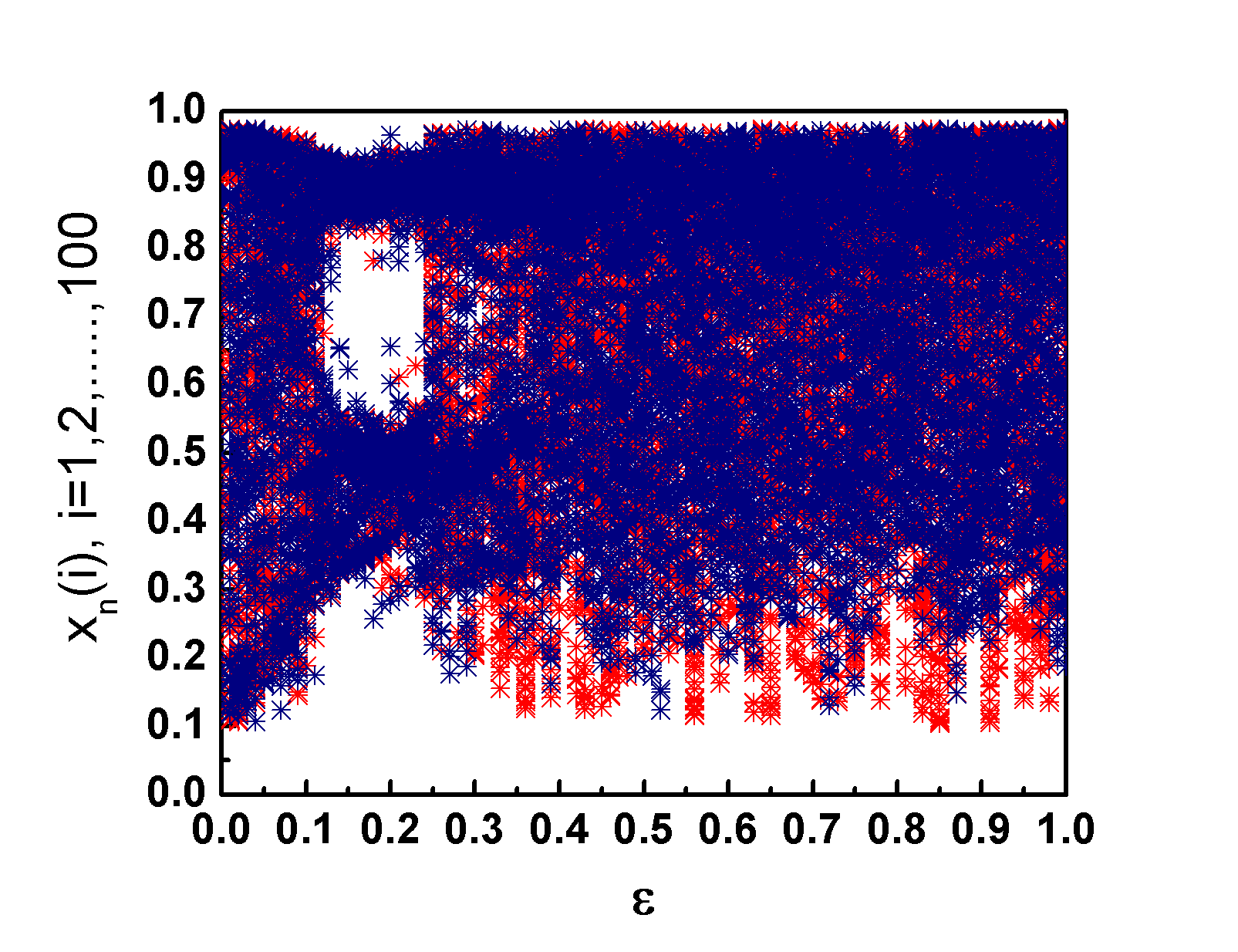}
\caption{\label{bifur} Bifurcation diagram of system. State variables $x_i$ 
are plotted as a function of coupling strength $\epsilon$ after
leaving sufficiently long transients for $p=0.0$ (red) and $p=0.1$ (blue).
We find the existence of 2-band attractor in PR region.
}
\end{figure}
$PR$ region is overall the same for all values of $p$. The 
nonlocal connections 
seem to make  only a marginal difference. The dynamical
origin of $PR$ region can be inferred 
from the bifurcation diagram. 
Fig.\ref{bifur} shows the 
bifurcation diagram for $p=0.0$ and $p=0.1$. 
In $PR$ region, we find a two-band attractor 
while in $NPR$ region the values of $x_i$ are spread over entire interval. 
Thus there is a attractor-widening crisis on either border.
For large $\epsilon$ and $p>0.7$, we observe spatial synchronization. 
This transition has been investigated in detail by several authors 
and we do not
study transition to chaotic synchronization in this work. We study the 
transition between $NPR-PR-NPR$, and explore the dynamics on critical 
lines. It is expected that 
the dynamics is slower for lower values of coupling. We denote the 
border on left side for $NPR-PR$ transition as border $I$ and the 
one on right side for $PR-NPR$ transition as border $II$. 

\begin{figure}[t]
\centering
\includegraphics[width=0.38\textwidth]{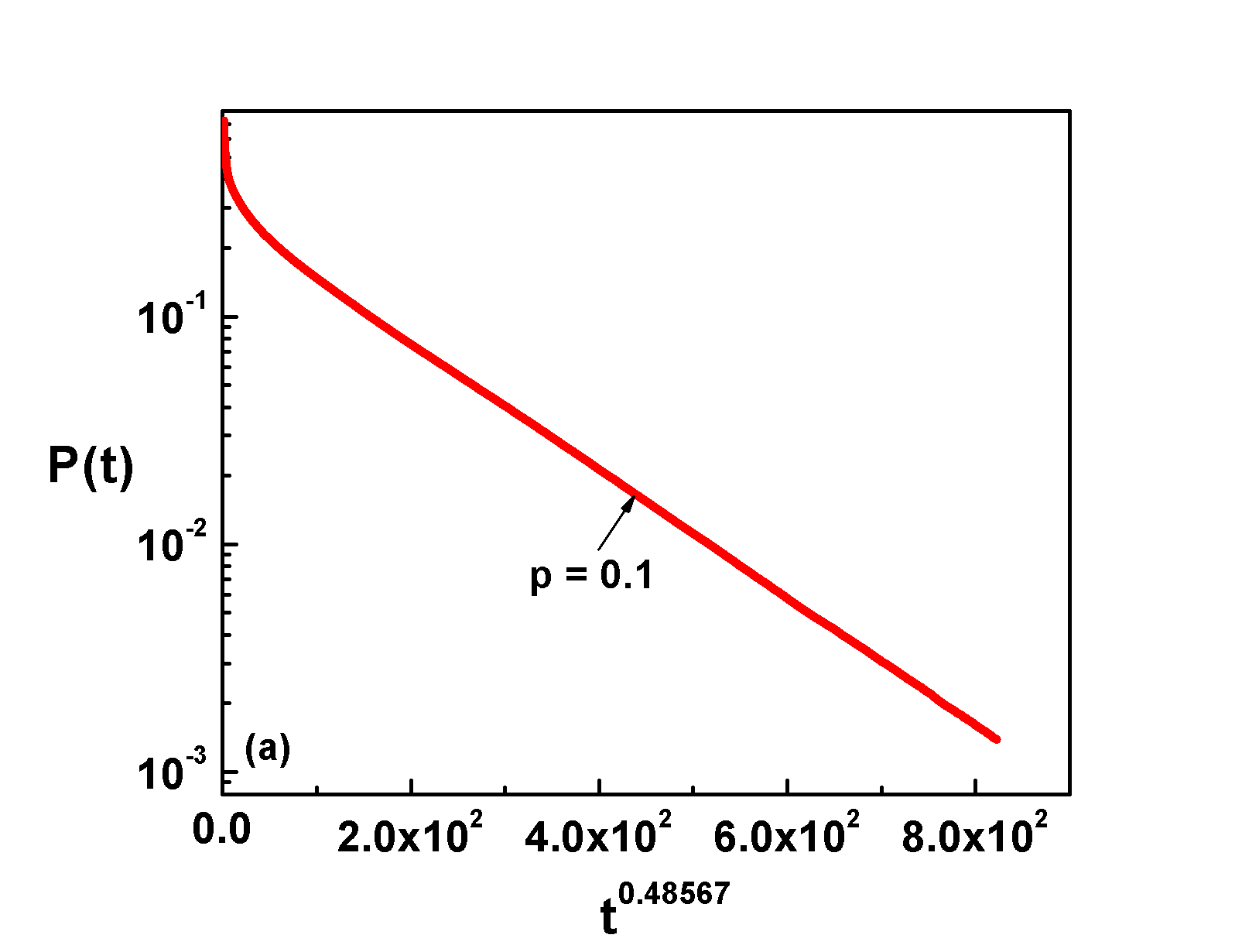}
\includegraphics[width=0.38\textwidth]{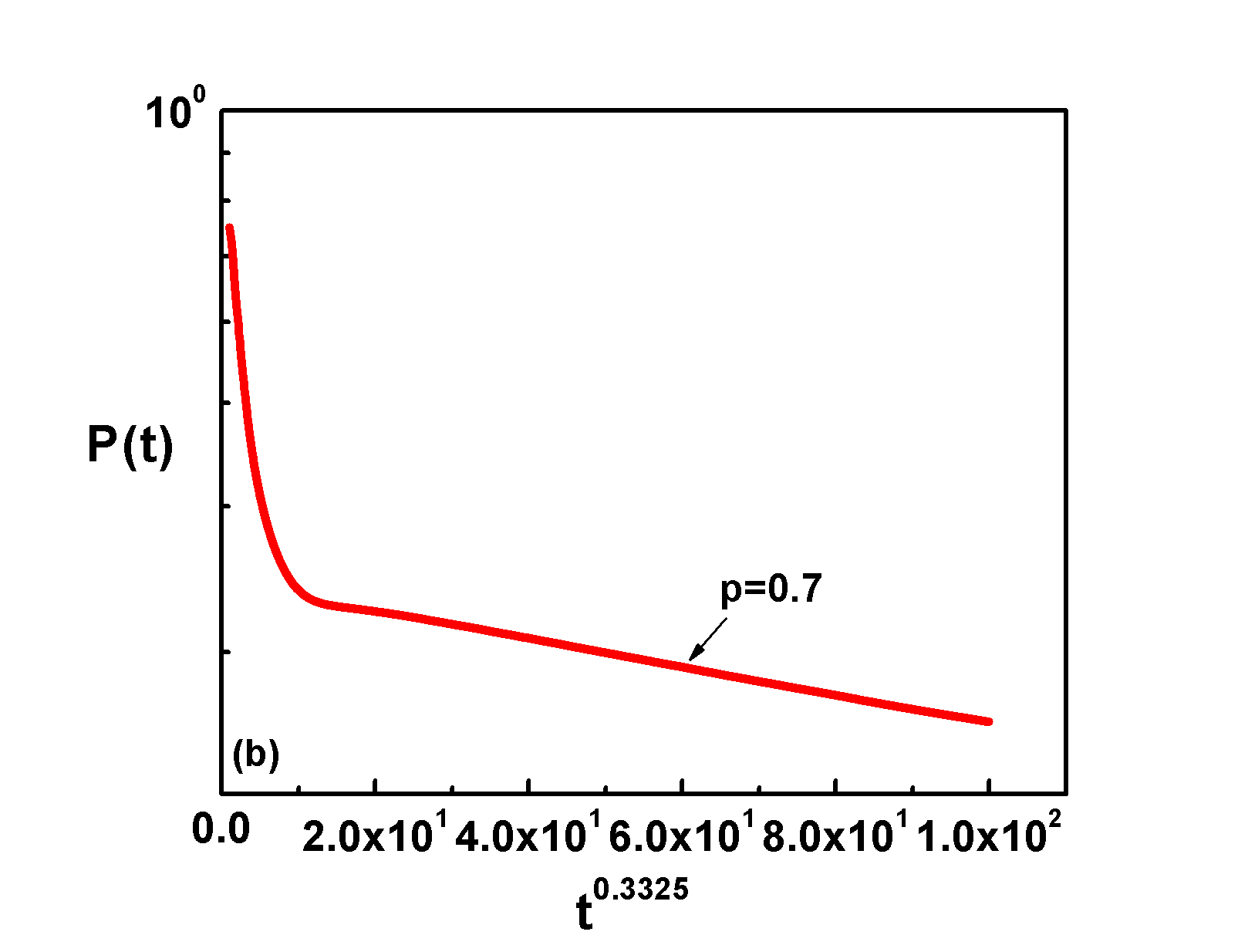}
\caption{\label{left}$P(t)$ is plotted as function of $t^{\beta}$ is plotted on border $I$ for $p=0.1$ and $p=0.7$ at critical coupling value $\epsilon=0.135$ and $\epsilon=0.152$ respectively, which clearly fits with a 
stretched exponential. Lattice size is $N=50000$. We averaged over 
$50$ initial conditions.}
\end{figure}

\begin{figure}[t]
\centering
\includegraphics[width=0.38\textwidth]{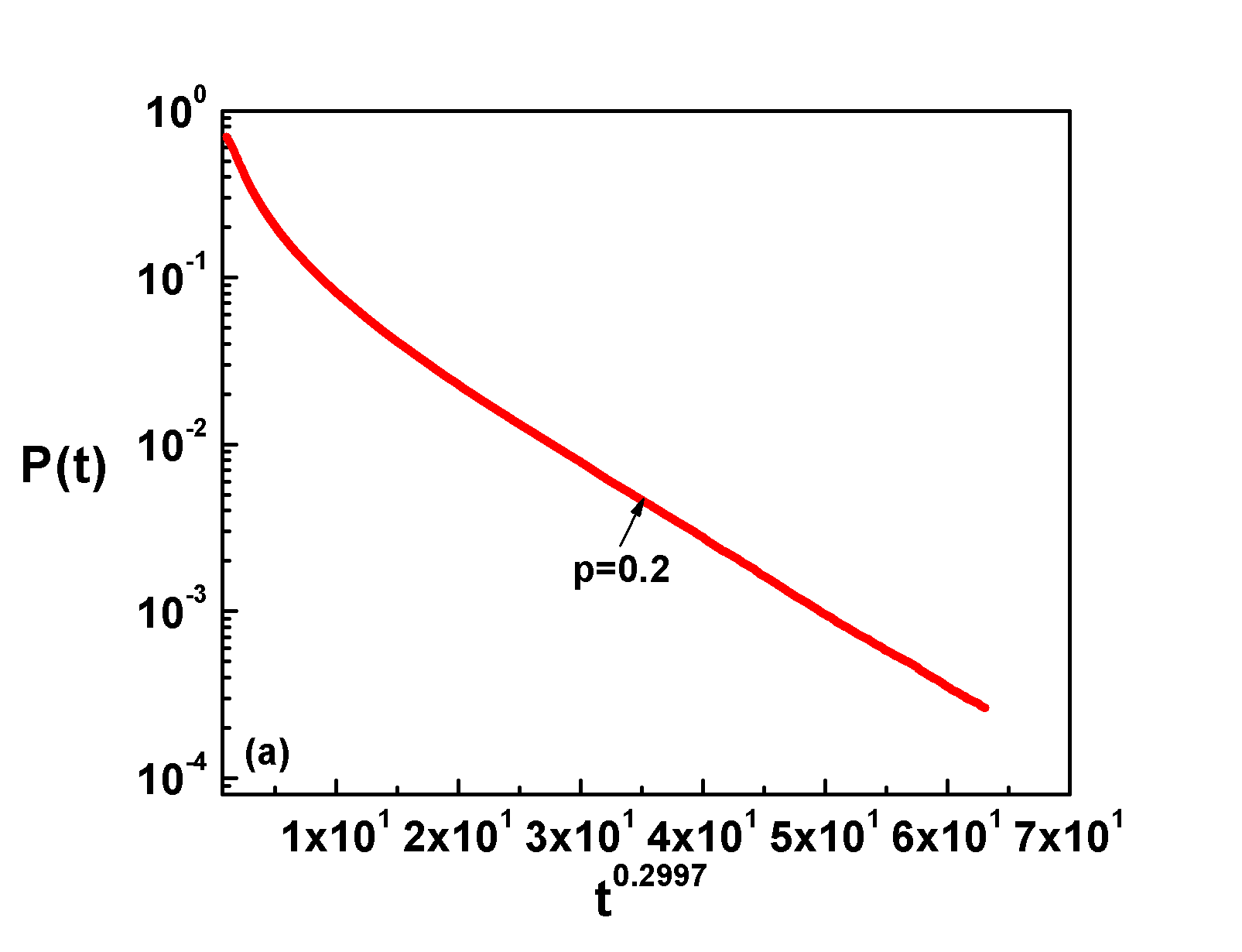}
\includegraphics[width=0.38\textwidth]{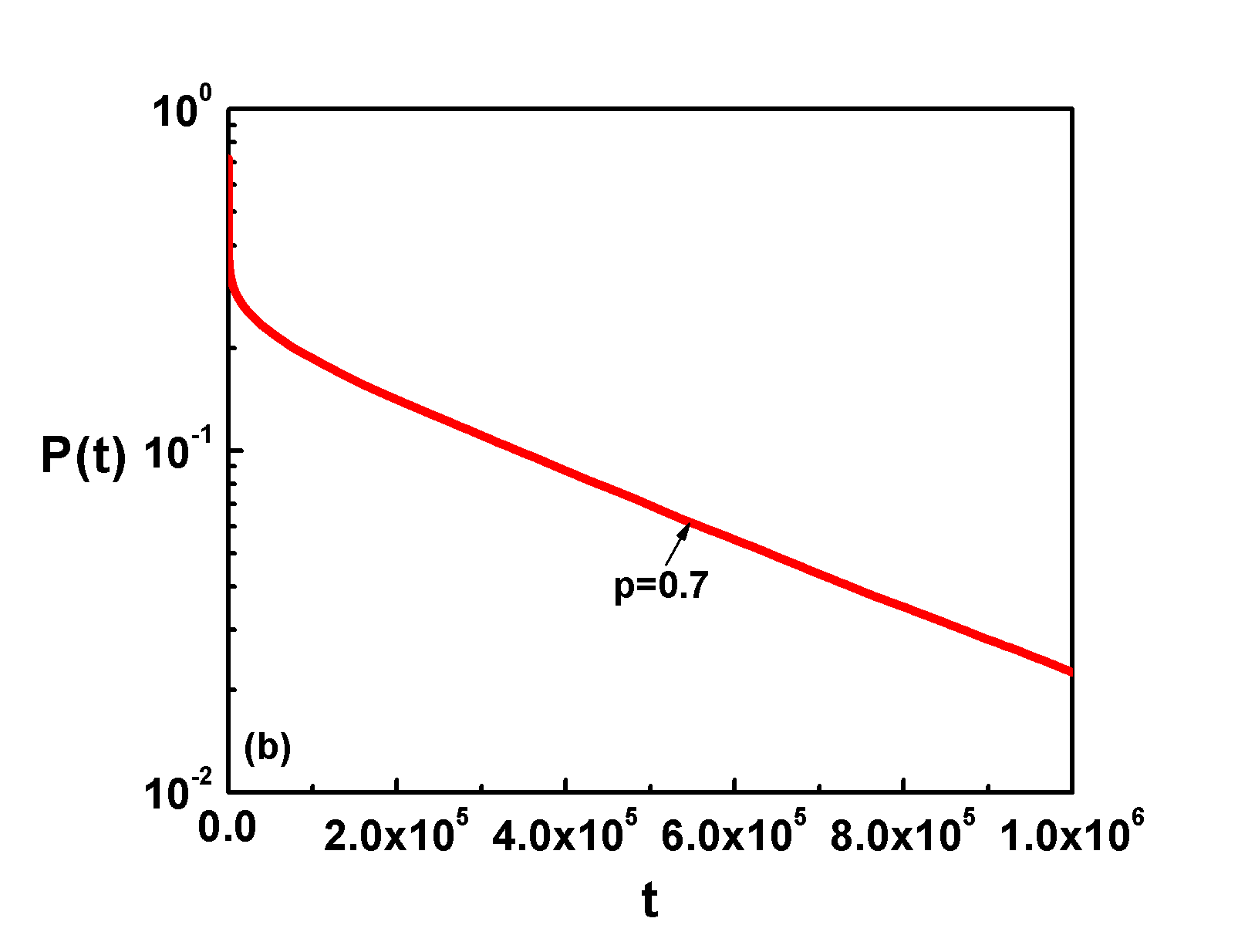}
\caption{\label{right}(a) Persistence $P(t)$ is plotted as 
a function of  $t^{\beta}$ on border $II$ with appropriate $\beta$. $\beta<1$ 
for smaller values of $p$ indicating stretched exponential decay
at critical value of $\epsilon$, (b) $\beta \rightarrow 1$ for higher 
values of $p$. Lattice size is $N = 5\times 10^4$. Data are averaged over $50$ 
initial conditions.}
\end{figure}

We investigate persistence and model it by a stretched exponential function
defined in eq.(1) which is an excellent fit at critical point for almost all points on both borders. It is defined by: 
\begin{equation}
f_{\beta}(\frac{t}{\tau_{0}}) = P(t) \sim \exp(-t^\beta)
\end{equation}
with $\tau_{0}=1$ and $0<\beta< 1$.  
Stretched exponential relaxation essentially is a signature of long-time confinement of dynamics to some regions in phase space.
On border $I$, we obtained stretched  exponential behavior
for all values of $p$. 
In Fig.\ref{left} $P(t)$ is plotted as function of $t^\beta$ on semi-log
scale for $p=0.1$ and $p=0.7$.   
On border $II$, we observe a stretched exponential behavior for small $p$
followed by  exponential for large values of $p$ (Fig.\ref{right}). 
Apart from visual fit, we have used standard programs such as Origin to
find the value of $\beta$ which gives the best fit. In literature,
$\beta$ is often approximated by closest rational and the values
we obtain are very close to simple rationals such as 
$\frac{1}{2}$,
$\frac{1}{3}$, 
$\frac{1}{4}$, 
$\frac{2}{5}$ or 1. However, we have plotted the unbiased estimators of 
$\beta$ in the figure.  
On border $I$, the $\beta$ is bounded by $\frac{1}{2}$ while it tends to 1
as $p\rightarrow 1$ on border $II$.
We also note that at early 
times,  persistence  drops faster than stretched
exponential. 
Initial conditions are chosen randomly from unit interval. Initially,
there are more negative spins since $x^*>1/2$. In a short time interval,
number of $+$ and $-$ spins becomes approximately equal. This leads 
to fast drop in persistence. 
Asymptotic behavior is very well described by stretched
exponential. For $p=0$ on border $II$, the value of $\beta$ obtained is very
close to zero, indicating that dynamics could even be slower. We obtain
equally good power law $1/t$ for persistence at this point. 

\begin{figure}[t]
\centering
\includegraphics[width = 0.5\textwidth]{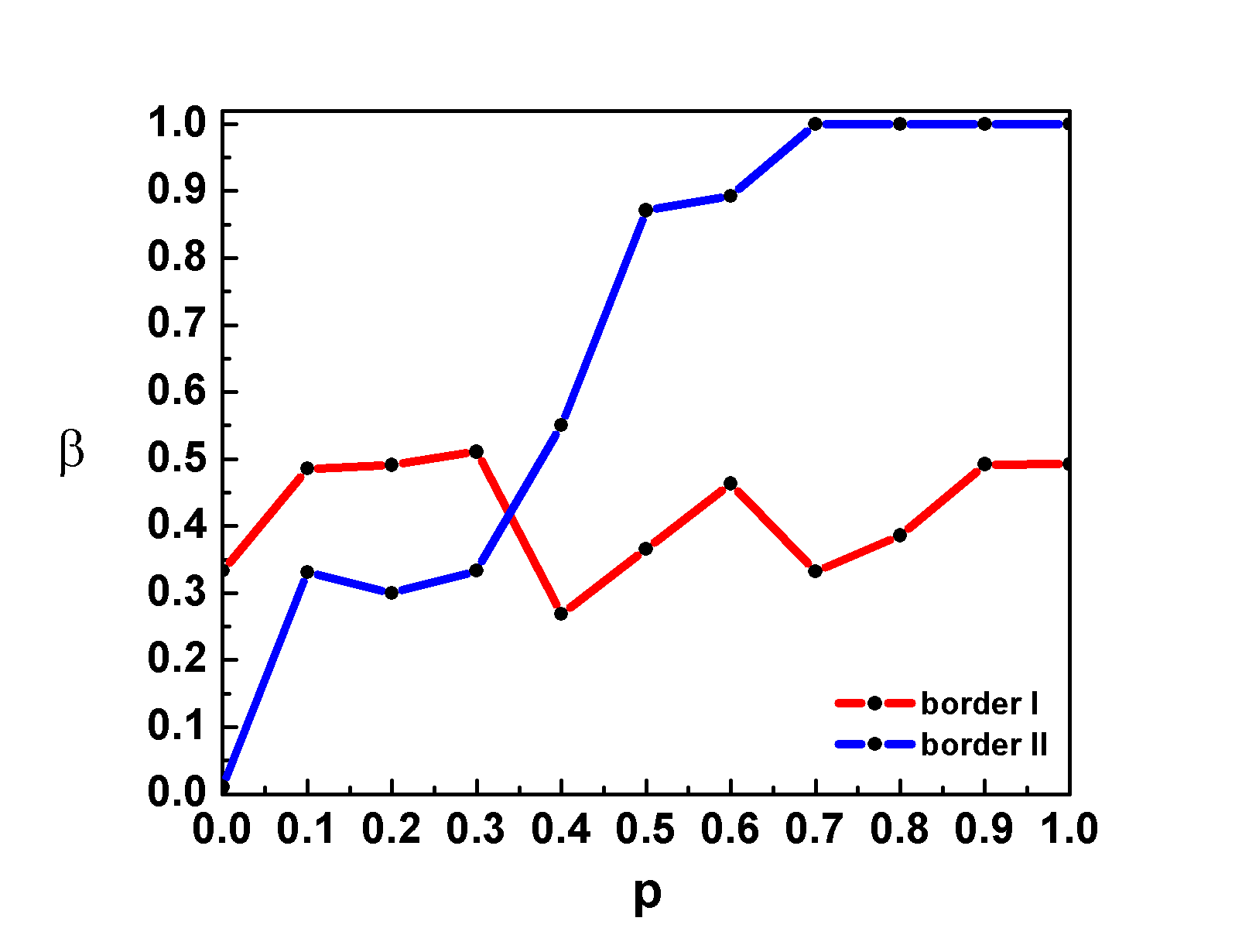}
\caption{\label{pvsb} Value of $\beta$ at critical point is plotted for different values of rewiring probability  $p$ on both borders.}
\end{figure}

Fig.\ref{pvsb} show the value of $\beta$ as a function of $p$ for 
border $I$ and $II$. 
As expected, the dynamics is slower for smaller values of couplings for 
both borders.
The nonzero asymptotic value of persistence indicates that 
some spin values never change from their initial conditions.

We introduce another 
parameter $F(t)$ to demonstrate the nonstationarity of dynamics.
$F(t)$ is the probability that spin value at time $2 t$ is different
from its value at time $2t-2$.
If this parameter goes to zero asymptotically,
the dynamics is completely frozen . If $F(t)$ and $P(t)$ are both nonzero 
asymptotically, it indicates partially frozen dynamics.
Here spins keep flipping, but some spins never flip. 
$F(t)$ is similar to hazard rate in survival statistics.
(The difference is that spin flips can occur multiple times unlike death.)
It decays as power-law
and goes to a very small nonzero value on border $I$. 
On border $II$, we observe a logarithmic decay followed by saturation. 
This indicates that though persistence decays as stretched exponential on
either border, detailed dynamical behavior is different. 
Within our computational limitations,
it indicates that spin state is not completely frozen but partially arrested
on both borders.

\begin{figure} 
\centering
\includegraphics[width=0.38\textwidth]{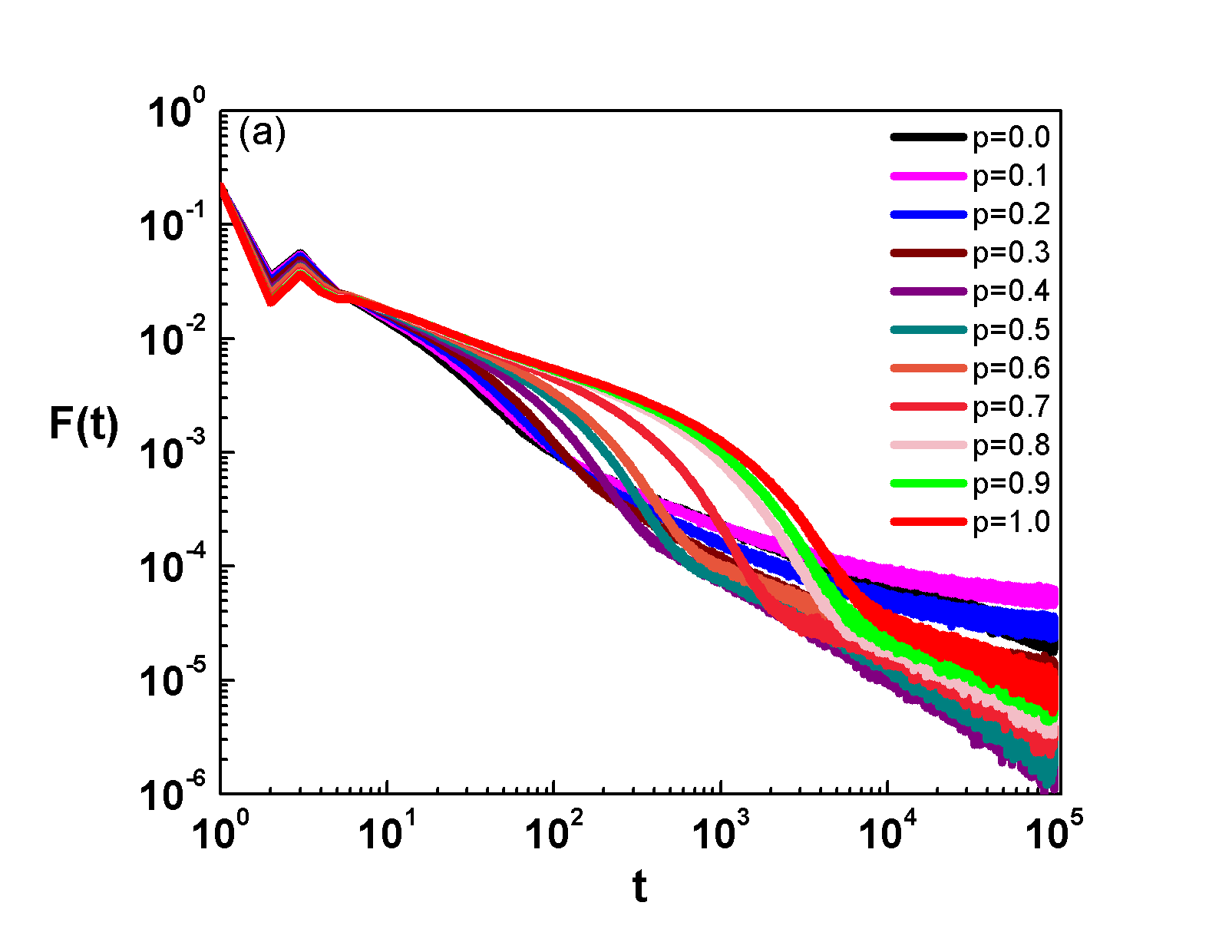}
\includegraphics[width=0.38\textwidth]{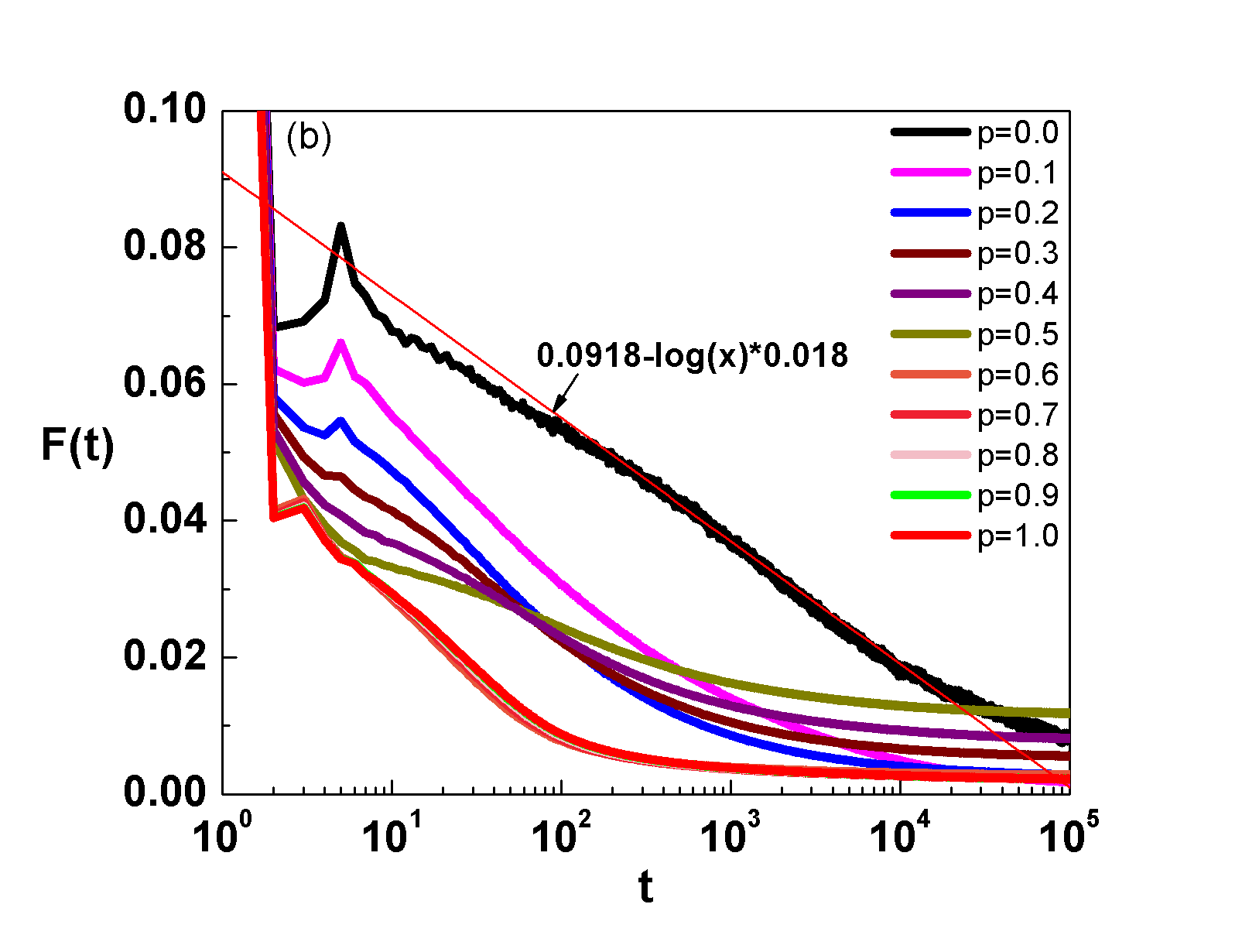}
\caption{\label{flip}  
(a) $F(t)$ decays as power-law on border $I$ for all $p$. 
(b) $F(t)$ decays as $\log(t)$ and saturates at critical coupling strength on border $II$. 
}
\end{figure}

\begin{figure}
\centering
\includegraphics[width=0.38\textwidth]{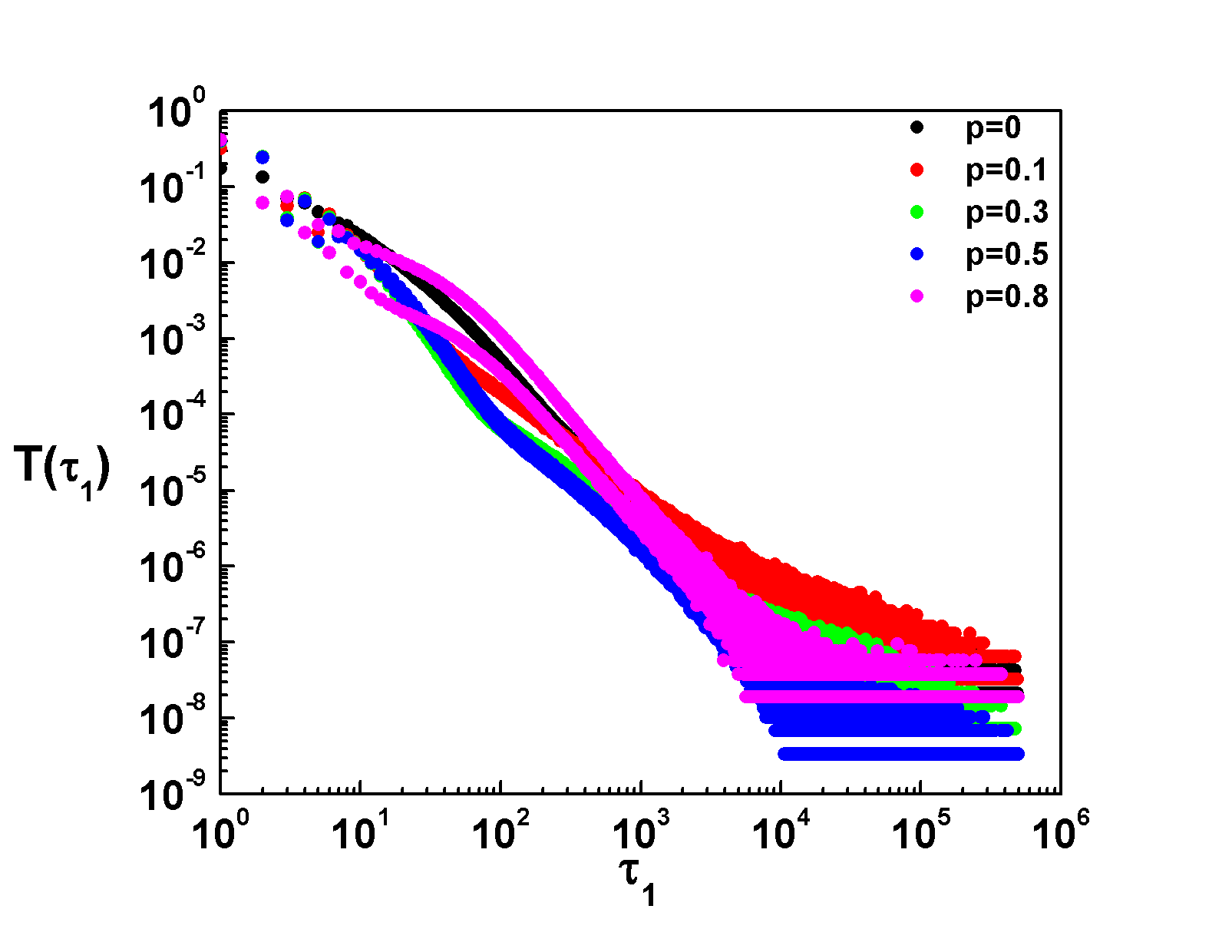}
\includegraphics[width=0.38\textwidth]{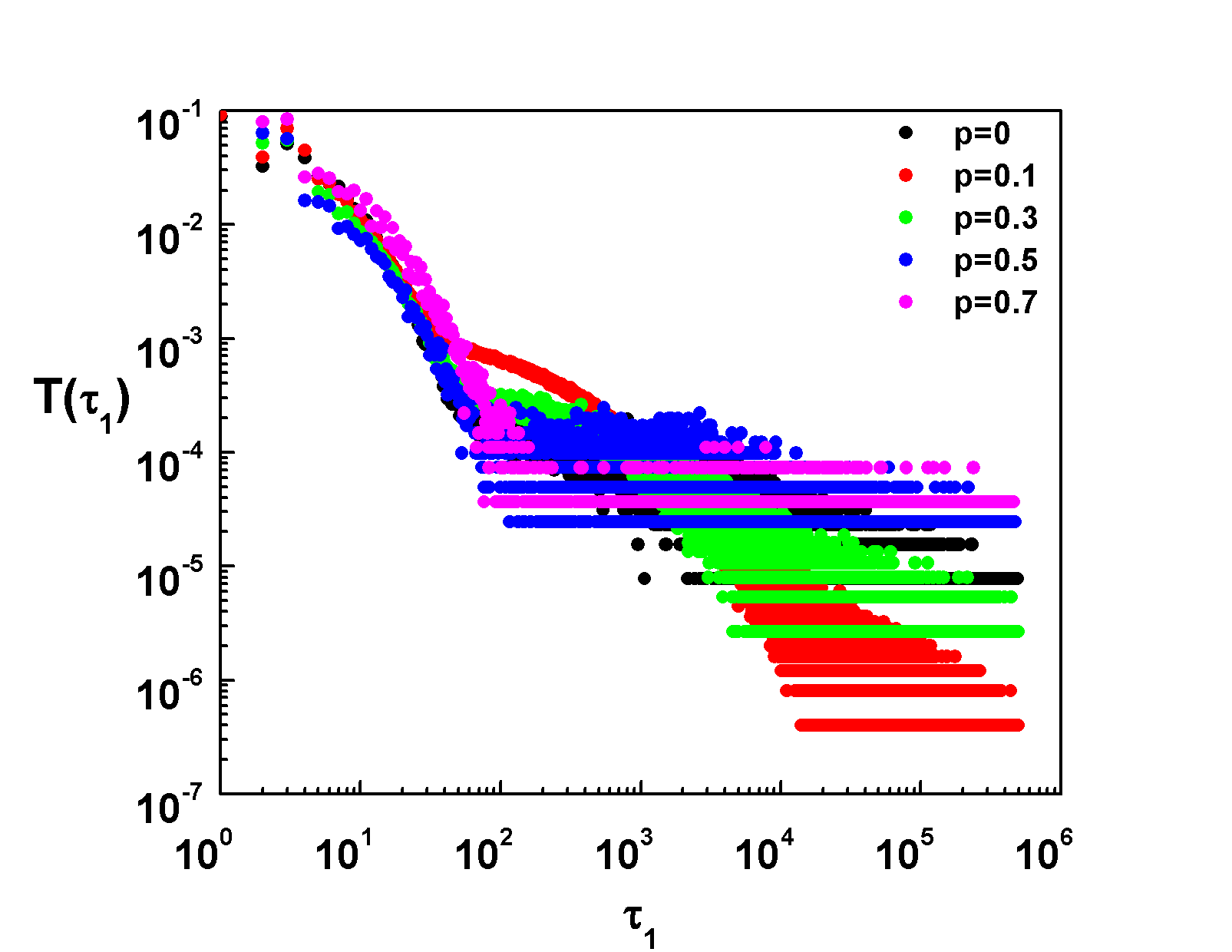}
\caption{\label{trap}Distribution of trap times for border $I$ and $II$, a) Tail of data fits to a power law on border $II$. b) Very long tailed
distribution is observed on border $I$.
}
\end{figure}

 
We also study trap time distribution. Trap time is the time period to 
which particular site stays in same state 
(not just initial state). 
If a site changes state at time $\tau$ and 
change it again at time $\tau+\tau_1$, it is trapped for $\tau_1$ period. 
We compute the probability distribution $T(\tau_1)$ for trap time $\tau_1$
The persistence is a generalization of first passage time
studies in stochastic systems while  
the trap times are similar to return times
in stochastic systems.
They usually do not obey same statistics. 
Since the system is highly nonstationary, this distribution is
expected to change in time. We wait for $10^7$ time-steps so that 
steady-state properties can  be observed. 
Fig.\ref{trap} shows that this distribution is well described by
power-law tail 
on border $II$. The exponent depends on value of $p$.
The behavior on border $I$ could not be 
approximated by 
a function with closed 
form. But it is clearly slower than exponential and longer
traps are more probable than border $II$ since the $F(t)$ is smaller
asymptotically.

\begin{figure}
\centering
\includegraphics[width=0.38\textwidth]{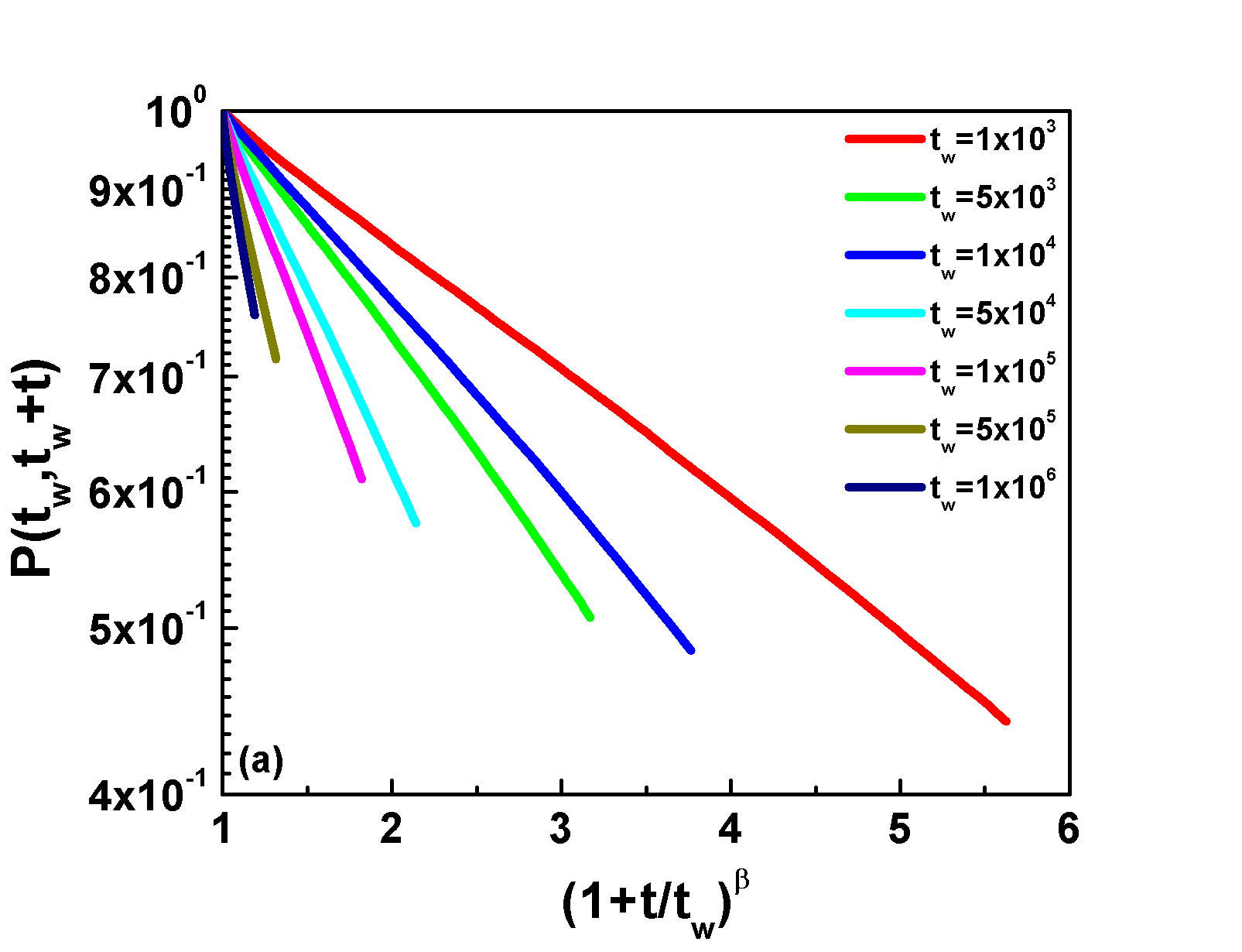}
\includegraphics[width=0.38\textwidth]{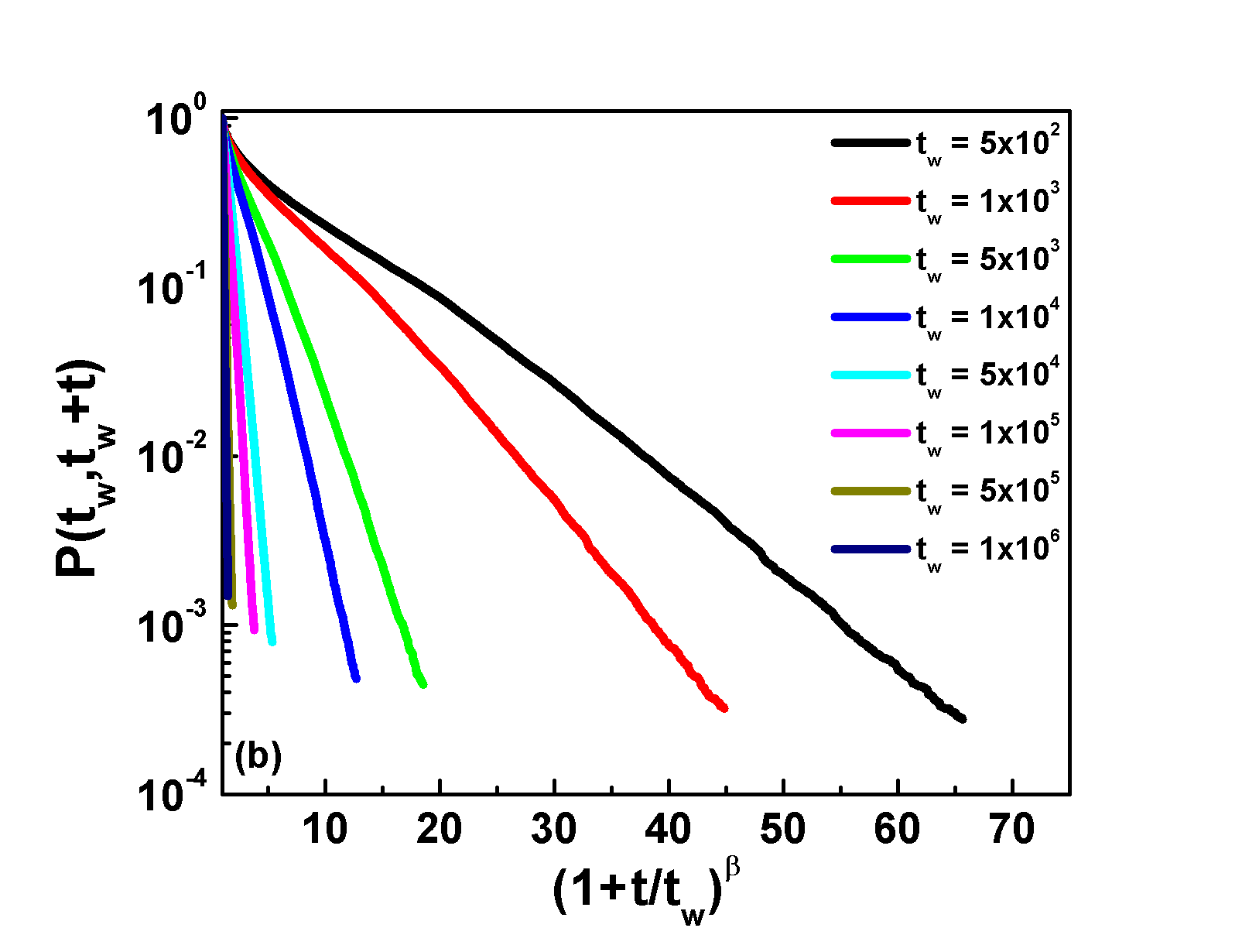}
\includegraphics[width=0.38\textwidth]{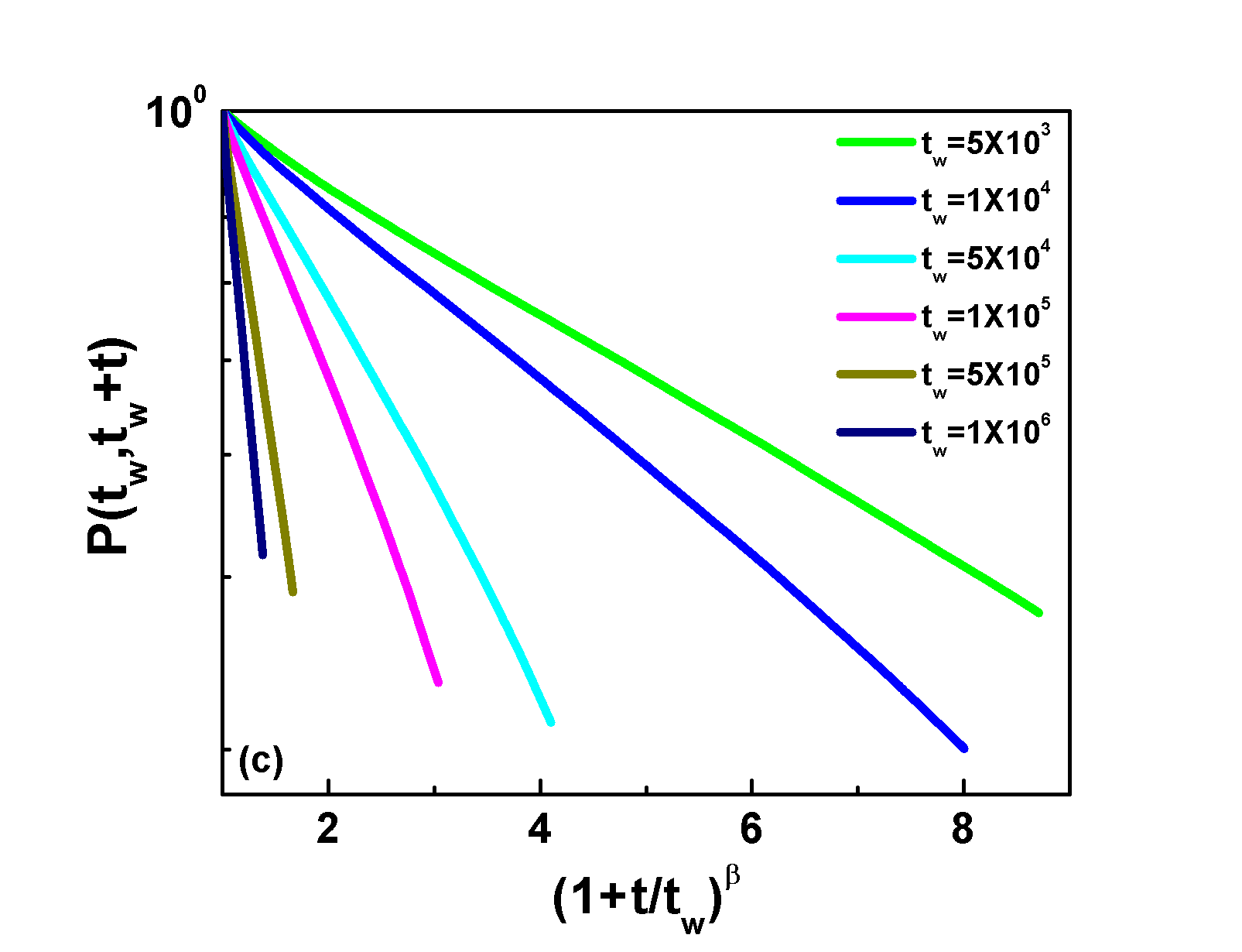}
\includegraphics[width=0.38\textwidth]{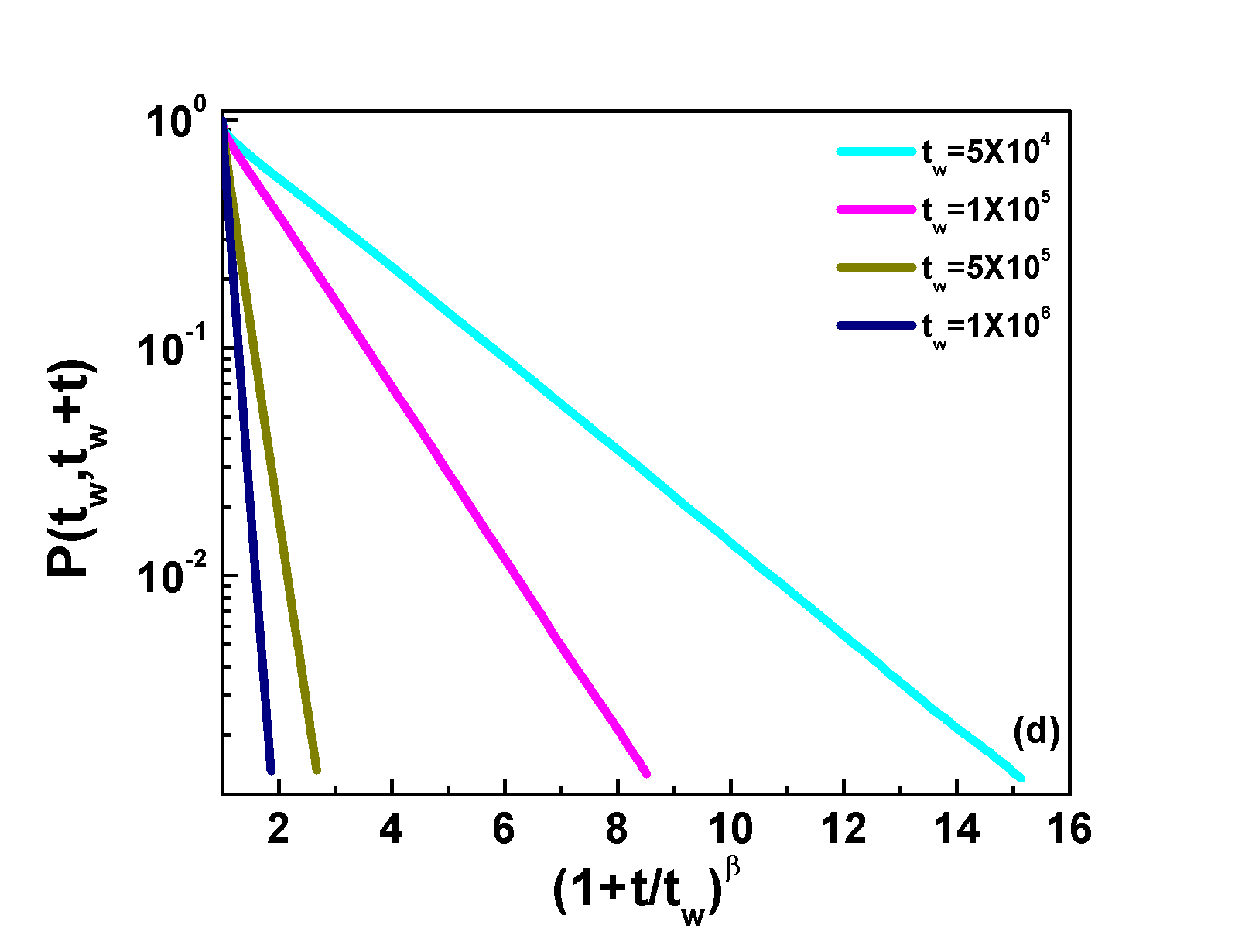}
\caption{\label{p02}  
 $P(t_{w},t_{w}+t)$ is plotted as function of $(1+t/t_{w})^\beta$ for 
a)$p=0.4$,
$\epsilon=0.145$ on border I using  $\beta = 0.26836$, 
b) $p=0.4$, $\epsilon=0.238$ 
for border $II$
using $\beta =0.55051$, 
c) 
$p=0.6$,  $\epsilon=0.147$ (border I) with $\beta = 0.46355$, and 
d)
$p=0.6$ and  
$\epsilon=0.1775$ (border II) using $\beta = 0.89267$ for various $t_w$. 
Clear straight lines are obtained on semi-log scale.
}
\end{figure}

\begin{figure}
\centering
\includegraphics[width=0.38\textwidth]{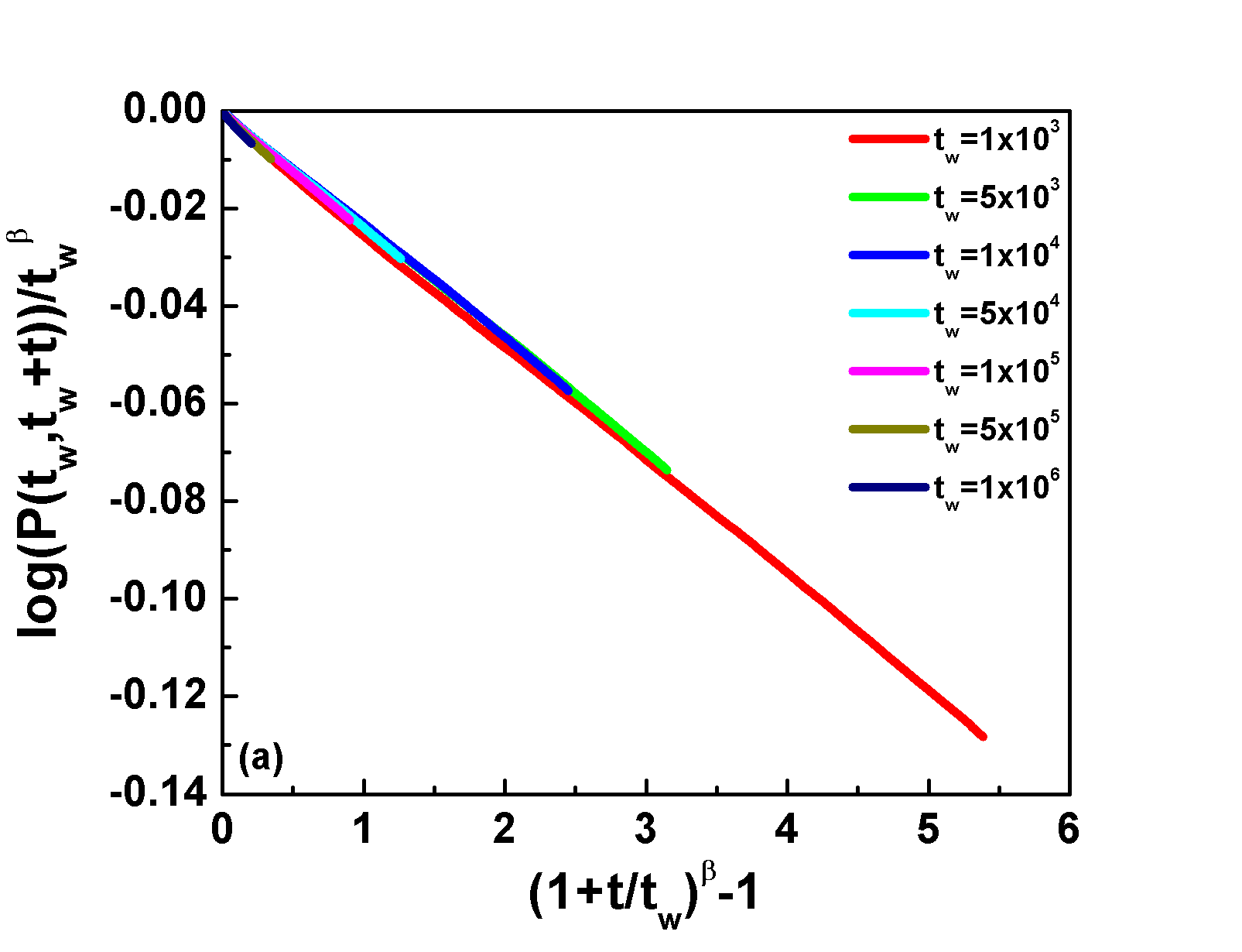}
\includegraphics[width=0.38\textwidth]{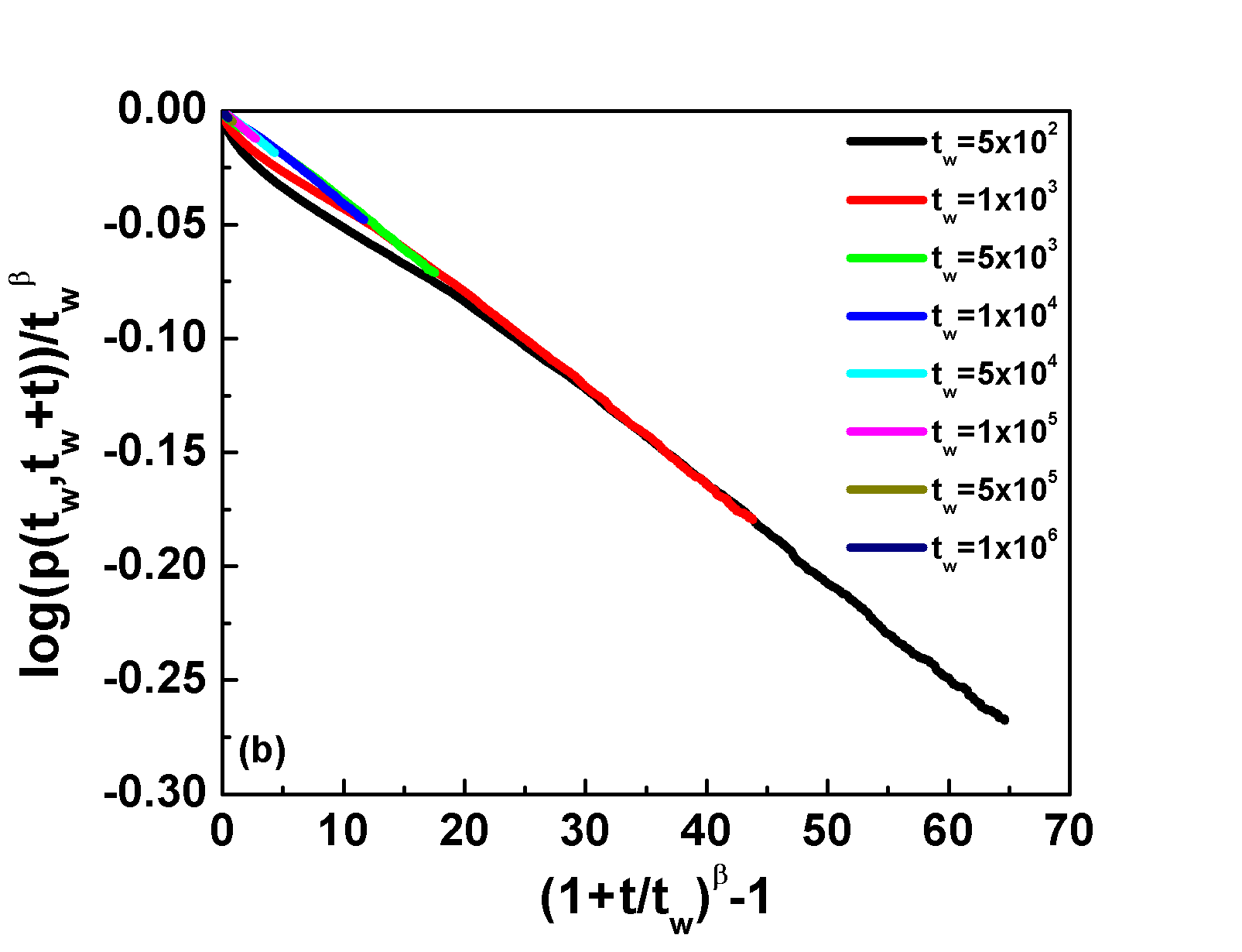}
\includegraphics[width=0.38\textwidth]{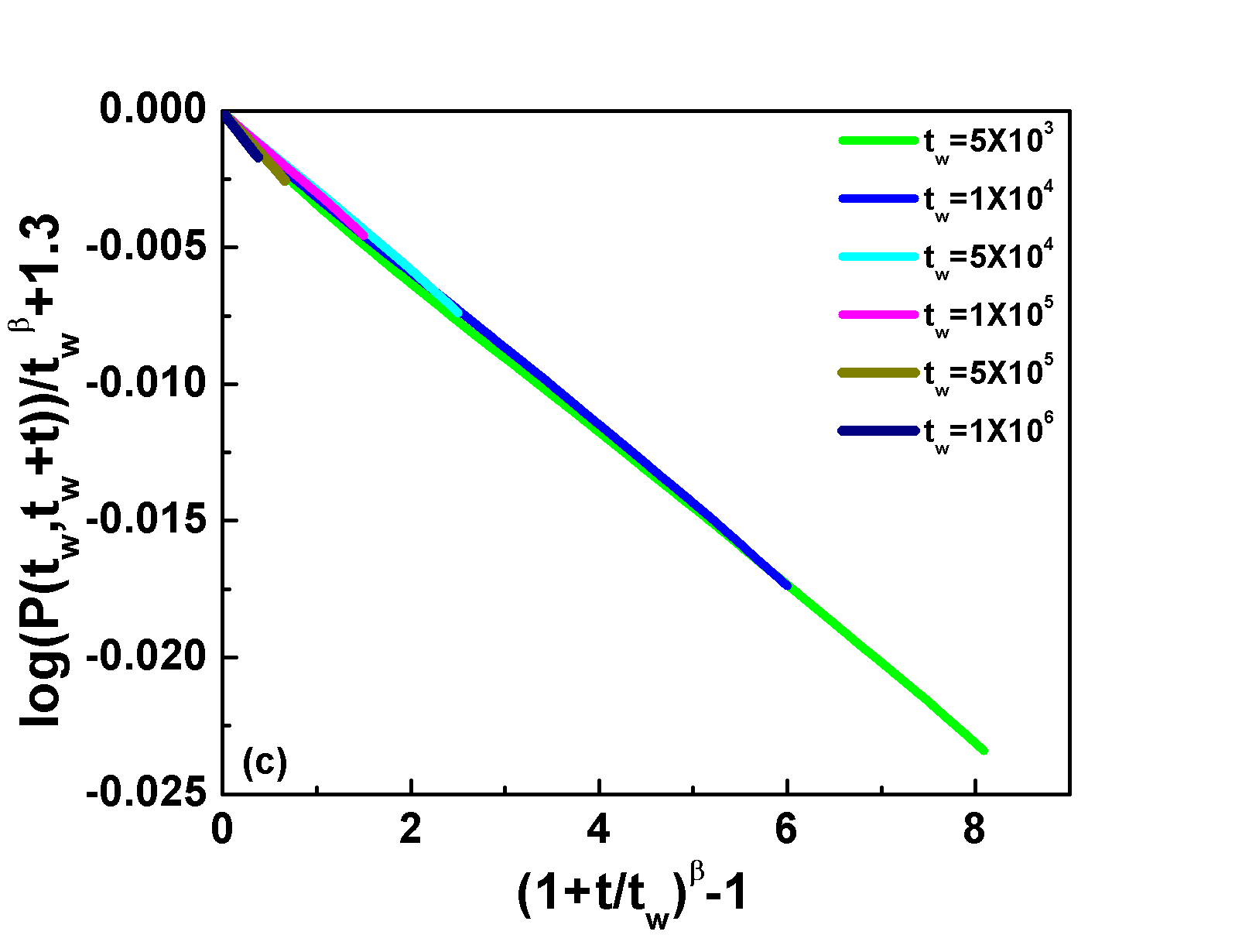}
\includegraphics[width=0.38\textwidth]{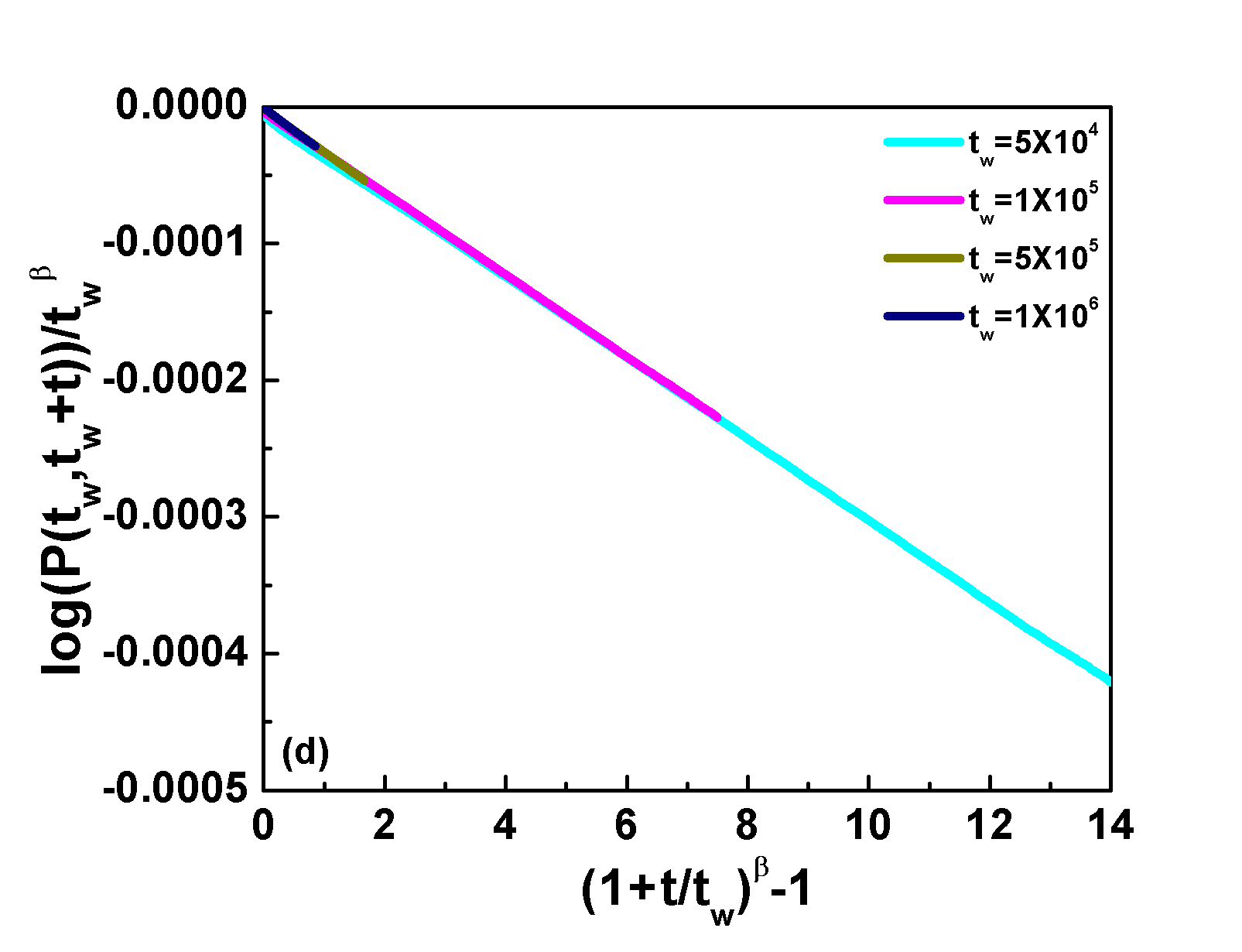}
\caption{\label{p04}  
Scaling collapse for data on Fig. \ref{p02}  a), \ref{p02}
b) \ref{p02}c) and \ref{p02}d) is 
obtained in a) b) c) and d)  according to Eq. \ref{sc}.
}
\end{figure}

In glasses,  
stretched exponential relaxation is often accompanied 
with aging. The flipping probability $F(t)$ clearly
depends on time elapsed in our system as well. 
The properties change in time. 
If we cool the system below glass transition temperature 
for time $t_w$ and bring it 
back to critical temperature, relaxation depends on the history of the 
system and time 
for which sample is cooled and is a function of $t/t_w$ in
general\cite{amir2012relaxations,shimer2010nonequilibrium}. If we 
carry out evolution inside the persistent 
regime for time $t_w$ and then bring system to critical coupling value
on either border, 
we do not observe any marked dependence on history of the system. 
However, if we allow system to equilibrate for time $t_w$ (waiting time) 
at any point on border $I$ or border $II$
and compute persistence after this time, we 
observe interesting results. 
Similar studies have been conducted in survival statistics. We may
study the survival of machine/ animal provided it has already
survived $t_w$ years. (These two studies are not exactly equivalent
since even sites which have not persisted till time $t_w$ are also
part of our statistics.)
If we assume that probability of persistence with respect
to initial conditions at $t_w$ time steps has same form as
survival probability for Weibull distribution after $t+t_w$, given it has
already survived till time $t_w$, the
following form for persistence after waiting time $t_w$ can be proposed:
\begin{equation}
P(t_w,t_{w}+t) =\frac{\exp(-r(t+t_{w})^\beta)}{\exp(-rt_{w}^\beta)}
\end{equation}
Thus
\begin{equation}
P(t_w,t_{w}+t) = - \exp(r t_{w}^\beta) \exp((1+t/t_{w})^\beta-1)
\end{equation} 
\begin{equation}\label{sc}
\frac{\log(P(t_w,t_{w}+t))}{t_{w}^\beta} = -r ((1+t/t_{w})^\beta-1)
\end{equation} 
The behavior  of $P(t_w,t_w+t)$ as a  function of $(1+t/t_w)^\beta$ on 
semi-log scale  is clearly linear along both borders even for smaller $t_w$.
This behavior is clearly observed in 
Fig. \ref{p02} and in Fig. \ref{p04}. 
(Still this fit is surprising due to approximation 
involved.)
This is another evidence that 
asymptotic behavior is indeed
described by stretched exponential. 
The validity of scaling form in eq.\ref{sc} 
depends on whether the persistence at time $t_w$  and $t+t_w$
is well defined by stretched exponential.
For very small and very large $t_w$'s  the assumption of stretched
exponential decay itself is  not valid either. There is a large drop
in persistence at early times.  Inevitable finite size effects as well
as error in finding critical point can affect long-time behavior. 
However, this scaling holds in most cases.
Fig. \ref{p04} shows
the scaling for data in Fig. \ref{p02}. 
To account for drop at early times, we propose a modified scaling law
\begin{equation}\label{sc1}
\frac{\log(P(t_w,t_{w}+t))}{t_{w}^\beta+C} = -r ((1+t/t_{w})^\beta-1)
\end{equation} 
where $C$ is a small constant. With this modification, above scaling holds
for a bigger range of $t_w$. 


There are few models of stretched exponential relaxation. Models such as
resonance energy transfer model are not very relevant for this work.
A model that could be of particular interest is random trap model.
Grassberger and Procaccia studied diffusion particle density in a 
$d$-dimensional medium (in which static traps are distributed randomly). 
Particle density decays as 
$\exp[-t^{d/d+2}]$ \cite{grassberger}. (In $1-d$, the 
decay will be $\exp[-t^{1/3}]$ and decay will be exponential 
as $d \rightarrow \infty$.)  For $p=0$ on border $I$, $\beta \sim \frac{1}{3}$
and for $p>0.6$ on border $II$, we obtain $\beta\sim 1$.
 SW network is an effectively infinite dimensional 
system because the characteristic distances between nodes grows as 
logarithm of system size\footnote{In a $d$-dimensional Cartesian network 
with $N$ sites ($N = L^d$ ), characteristic length scale  is $L = N^{1/d}$ 
and since logarithm is slower than power-law, it is effectively infinite 
dimensional network.}. Therefore it is expected that systems on a small-world 
network shows mean-field type behavior. Several equilibrium systems on a 
small-world network display behavior in the universality class as mean-field 
models for infinitesimal small values of $p$. However in non-equilibrium case, 
there are several examples in which mean-field behavior is not reached 
even for $p=1$ \cite{ashwini, ssinha, campos}.  
On border $II$, we observe mean-field type behavior for 
large rewiring probability. Persistence decays exponentially for $p>0.6$. 
On border $I$, we do not find such crossover. High coupling strength 
and small-world effect (involvement of more random connections due to 
large $p$) might be a reason of crossover on border $II$. 

\begin{figure}
\centering
\includegraphics[width=0.38\textwidth]{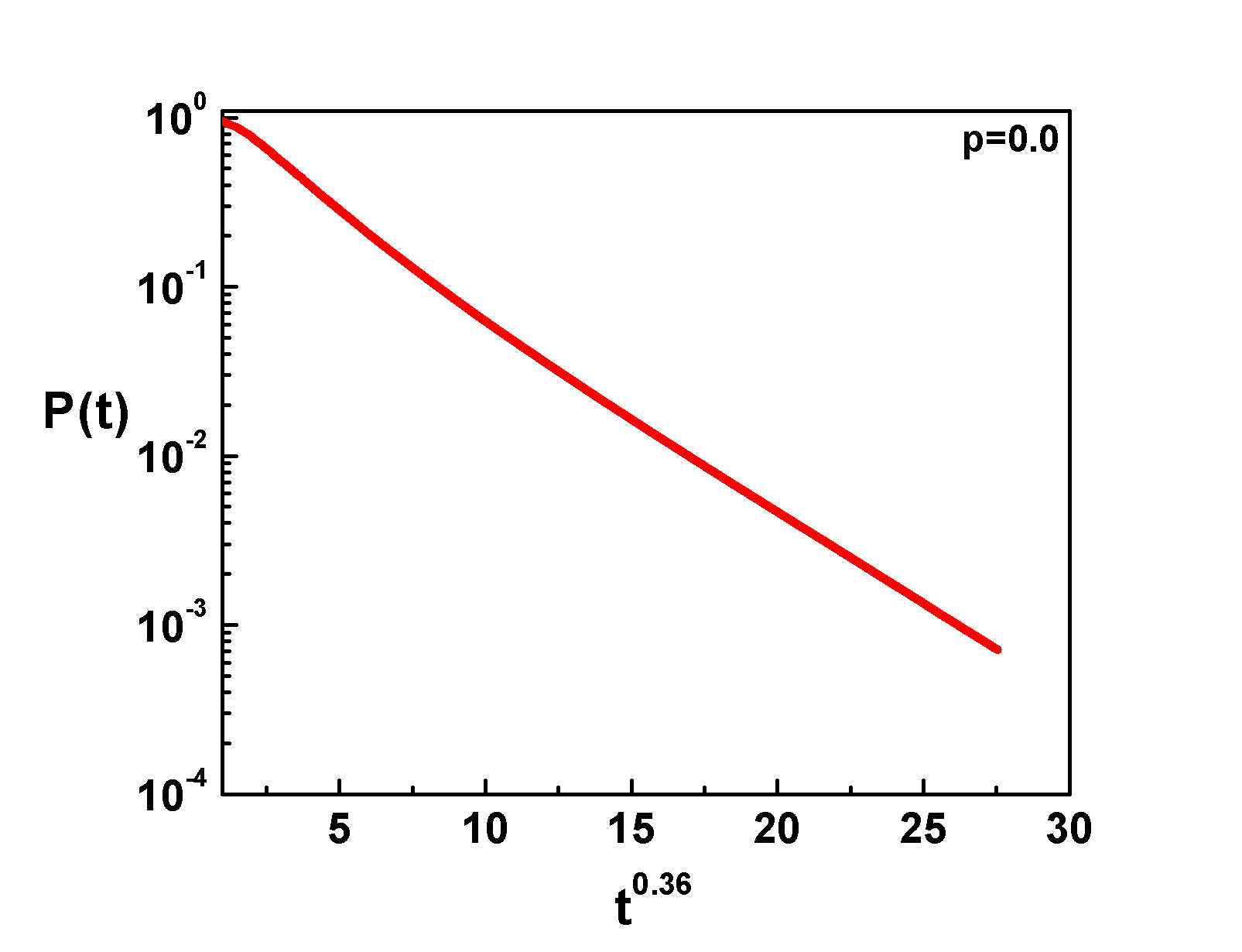}
\includegraphics[width=0.38\textwidth]{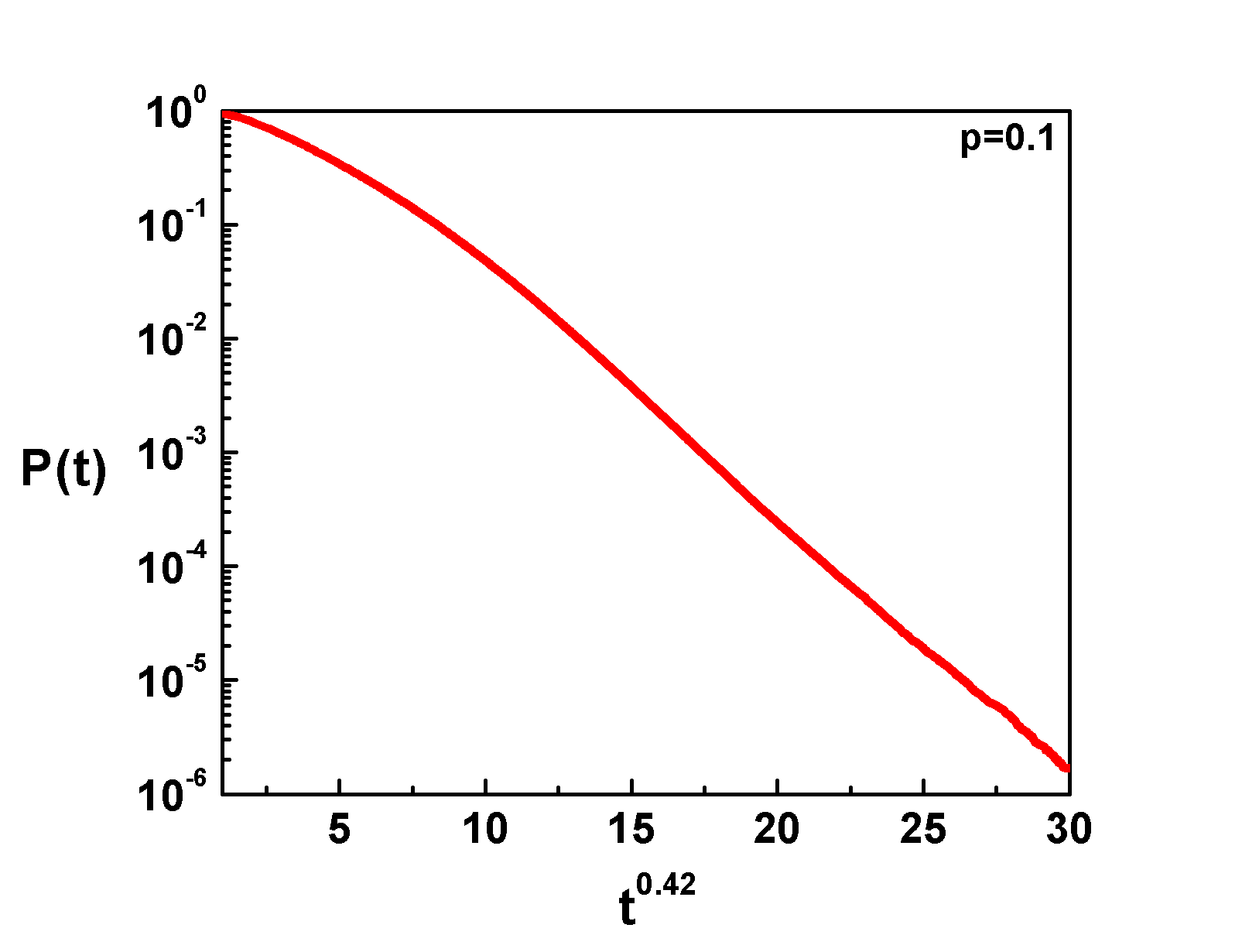}
\includegraphics[width=0.38\textwidth]{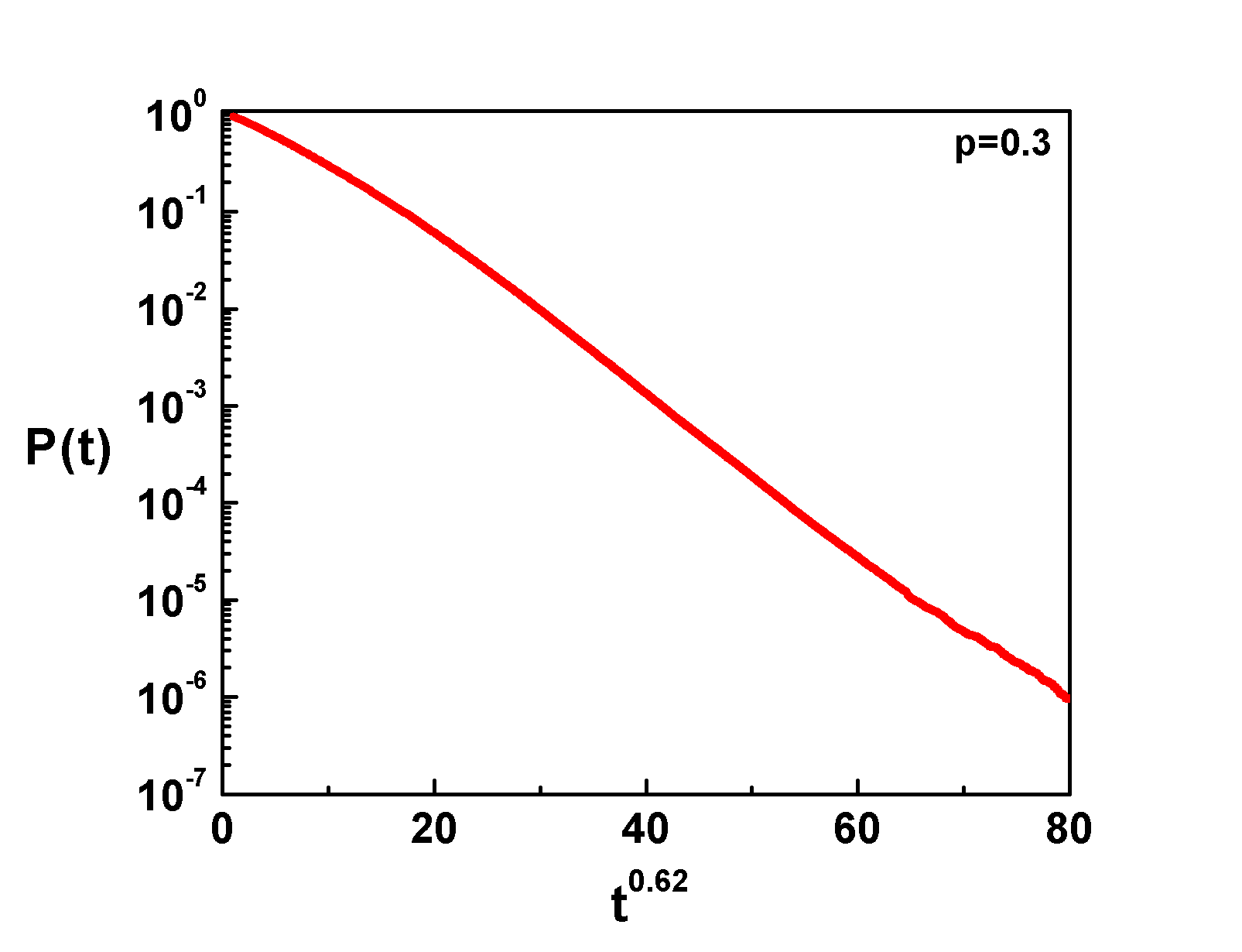}
\includegraphics[width=0.38\textwidth]{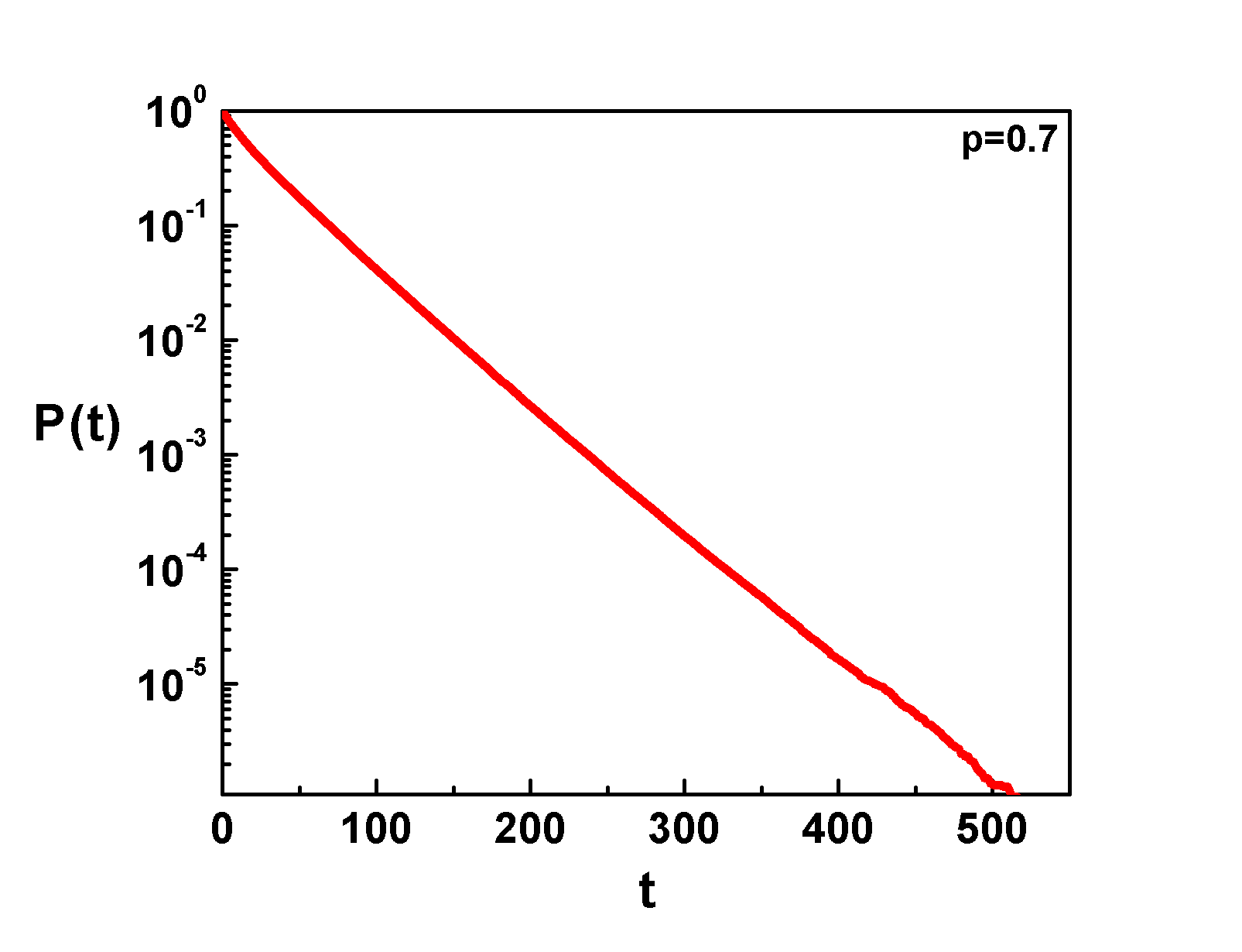}
\caption{\label{ising}  
 $P(t)$ is plotted as function of $t^\beta$ on semi-log scale
for $p = 0$, $p=0.1$, $p=0.3$ and $p=0.7$. It can be observed that data
can be fitted by stretched exponential. The value of $\beta$ increases
from $\beta=0.36$ for $p=0$ to 1 as $p\rightarrow 1$.
}
\end{figure}

We propose another possible model for stretched exponential relaxation.
Let us consider $N$ spins which are assigned value 1 or -1 
initially. Each spin is coupled to two nearest
neighbors on either side with antiferromagnetic coupling. To generate SW
network, we rewire the links with probability $p$ as in previous section.
All sites are updated in parallel with following rule.
a) If sum of spins of all neighbors is positive, the spin value is set to -1.
b) If sum of spins of all neighbors is negative, the spin value is set to 1.
b) If sum of spins of all neighbors is zero, the spin value is changed
with probability $q$. We set $q=0.05$. 
With these rules, we obtain stretched exponential relaxation. As shown in fig.\ref{ising}, the exponent 
gradually increases from $0.36$ to 1 with increasing $p$.
 
\section{Conclusion}
\label{sec4}
We study coupled logistic map on a small-world 
networks and explore the phase diagram 
for persistence 
in $\epsilon-p$ space.
The persistent region is a bounded by two critical lines
$I$ and $II$.
Persistence on both borders wows stretched exponential decay with
crossover
to exponential on border $II$ for large $p$.
We also  obtain scaling
form for decay of persistence after waiting for time
$t_w$. It has the same form on both the borders. 
However, other quantifiers flip probability $F(t)$ and
trap time distribution show different behavior along both borders.
Trap time distribution has power law decay in time along border $II$ while
no such behavior is obtained on border $I$. On the other hand, $F(t)$ shows
power-law decay on border $I$ and logarithmic decay on border $II$.
We believe that exchange frustration is key to understanding this behavior
and  present a Ising-type model to explain it.

%
\section{Acknowledgement}
AVM acknowledges DST WOS-A scheme for financial support 
and  A. Banpurkar and S. Barve for helpful discussion. 
PMG thanks RTMNU and DST research schemes for financial assistance.

\section*{References}
 


\end{document}